\documentclass[a4paper,fleqn,usenatbib]{mnras}

\usepackage{newtxtext,newtxmath}

\usepackage[T1]{fontenc}
\usepackage{ae,aecompl}
\usepackage[usenames,dvipsnames,svgnames,table]{xcolor}

\usepackage{graphicx}	
\usepackage{amsmath}	
\usepackage{amssymb}	
\usepackage{longtable}

\defcitealias{gon15}{Paper~I}
\defcitealias{dor16a}{Paper~II}
\defcitealias{dor16b}{Paper~III}

\title[LTE $T_{\rm eff}$ scale for RSGs]{An LTE effective temperature scale for red supergiants in the Magellanic clouds}

\author[H. M. Tabernero et al.]{
H. M. Tabernero,$^{1}$\thanks{E-mail: htabernero@ua.es}
R. Dorda,$^{1}$
I. Negueruela$^{1}$
and
C. Gonz\'alez-Fern\'andez$^{1,2}$
\\
$^{1}$Departamento de F\'{\i}sica, Ingenier\'{\i}a de Sistemas y Teor\'{\i}a de la Se\~nal, Universidad de Alicante,
\\ Carretera de San Vicente del Raspeig, E03690 Alicante, Spain\\
$^{2}$Institute of Astronomy, University of Cambridge, Madingley Road, Cambridge CB3 0HA, United Kingdom\\
}

\date{Accepted XXX. Received YYY; in original form ZZZ}

\pubyear{2017}

\begin{document}
\label{firstpage}
\pagerange{\pageref{firstpage}--\pageref{lastpage}}
\maketitle

\begin{abstract}
We present a self-consistent study of cool supergiants (CSGs) belonging to the Magellanic clouds. We calculated stellar atmospheric parameters using LTE KURUCZ and MARCS atmospheric models for more than 400 individual targets by fitting a careful selection of weak metallic lines. We explore the existence of a $T_{\rm eff}$ scale and its implications in two different metallicity environments (each Magellanic cloud). Critical and in-depth tests have been performed to assess the reliability of our stellar parameters (i.e.\ internal error budget, NLTE systematics). In addition, several Montercarlo tests have been carried out to infer the significance of the $T_{\rm eff}$ scale found.  Our findings point towards a unique $T_{\rm eff}$ scale that seems to be independent of the environment.
\end{abstract}




\section{Introduction}
\label{intro}

Red supergiants (RSGs) are moderately high-mass stars (between $\sim10$ and $40\:$M$_{\odot}$) that have evolved to the cool side of the Hertzsprung-Russell diagram. The RSG phase represents the last (or at least one of the latest) step in their evolution. It covers only a small fraction of their lifespan, being a powerful constraint to test evolutionary tracks \citep[see][]{eks2013}. Moreover, RSGs are the main progenitors of core-collapse supernovae (mostly Type IIp, but probably also other subtypes). For these reasons, the physical characterisation of these stars is fundamental in the understanding of their evolution. In this work we focus on the effective temperature ($T_{\rm eff}$) of RSGs, which is an open question topic still today.

The temperature scale is the relation between the spectral type (SpT), which is a morphological classification, and $T_{\rm eff}$. Classically, SpT classification of RSGs has been considered mostly related to the $T_{\rm eff}$, as in other types of stars. However, a new interpretation of SpT has recently been proposed by \cite{dav2013}, and in this work, we revisit the $T_{\rm eff}$ scale to test the new scenario with a statistically significant sample of RSGs. This scale is important even today, because it is easier to obtain a SpT classification than a temperature in many cases: SpT can be derived from mid to low resolution spectra and it does not require a S/N as high as a $T_{\rm eff}$ calculation. Moreover, not all groups have access to the resources necessary to derive precise temperatures. Although the SpT cannot supersede an $T_{\rm eff}$ calculation, knowing the relation between SpT and $T_{\rm eff}$ can be useful for many groups for the interpretation of their observations, especially in the case of large samples of RSGs.

Observational studies done in the last decades of the 20$^{\rm th}$ century \citep[e.g.][]{lee1970,hum1984} presented relatively cool $T_{\rm eff}$ scales for RSGs, spanning from $4\,300\:$K at K0 to $2\,800\:$K at M5. Two decades later, \cite{mas2003b} derived a slightly different scale (with lower temperatures for K subtypes and higher ones for M subtypes). However, as they discussed, the $T_{\rm eff}$ scale they obtained is too cool when compared to contemporary evolutionary tracks \citep{mey2000}. A few years later, \cite{lev2005} revisited this topic. They employed synthetic spectra, generated using MARCS atmospheric models, against spectrophotometric observations of RSGs from the Milky Way (MW), covering the range from $4\,000$ to $9\,500\:$\AA{}. Their $T_{\rm eff}$ scale for MW RSGs, based on the overall shape of the flux-calibrated spectra and the depth of TiO bands, resulted in a better agreement with \cite{mey2000}, since they derived warmer temperatures. Their temperature scale was also flatter than those in previous works, spanning from $4\,100\:$K at K1 to $3\,450\:$K at M5.

\cite{lev2006} also studied the RSGs from both the Large Magellanic Cloud (LMC) and the Small Magellanic Cloud (SMC). They used the same method as for the MW \citep{lev2005} to obtain a $T_{\rm eff}$ scale for each Magellanic Cloud (MC). Although their $T_{\rm eff}$'s were closer to those present in evolutionary tracks than those of previous works on these galaxies \citep{mas2003b}, they still presented some disagreement, especially for the SMC. RSGs in both galaxies span a similar $T_{\rm eff}$ range: they range from $\sim4\,200\:$K at K1 to $3\,475\:$K at M2 for the SMC, while for the LMC the range was from $\sim4\,300\:$K at K1 to $3\,450\:$K at M4. Despite their similar values, typical temperatures of RSGs in each galaxy are substantially different, because each population has a different typical SpT. \cite{hum1979b} had already found that the average SpT of a given population of RSGs has some dependence on its average metallicity, with later spectral types being found in higher metallicity environments. This effect has been widely confirmed within different galaxies \citep{eli1985,mas2003b,lev2012,dor16a}, and also along the galactic plane of a given galaxy \citep[M33;][]{dro2012}. In consequence, the results of \cite{lev2006} indicate that RSGs from the SMC have a typical $T_{\rm eff}$ of $\sim3\,920\:$K, while it is $\sim3\,705\:$K in the case of the LMC.

Against these results, recently \cite{dav2013} discussed the limitations of the methodology used by \cite{lev2005,lev2006}. They analysed a small spectrophotometric sample of stars from both MCs, obtaining their $T_{\rm eff}$'s by using different methods. Firstly, they employed the same procedure as \cite{lev2006}, by performing a global fit of their spectra to synthetic spectra generated using MARCS stellar atmospheric models. Since TiO bands dominate the appearance of RSG spectra in the optical region, they presented this method as a TiO scale. They also derived $T_{\rm eff}$'s by fitting the optical and infrared (IR) spectral energy distribution (SED). Finally, they used the flux integration method (FIM) as a constraint for the two other methods. They found that the $T_{\rm eff}$'s  obtained by fitting MARCS-based synthetic spectra are significantly cooler than those obtained through the SED. They argued that the SED $T_{\rm eff}$'s seem rather more reliable because of three strong arguments: i) MARCS $T_{\rm eff}$'s were significantly cooler than those obtained through the FIM at the lowest interstellar reddening; ii) although the $T_{\rm eff}$'s obtained through MARCS synthetic fitting reproduce well the TiO bands in the optical range, they overpredict the IR flux; and iii) they found no correlation between the reddening obtained through the TiO method and the diffuse interstellar bands measured. Moreover, their SED $T_{\rm eff}$'s agree with Geneva evolutionary tracks. They argued that the discrepancies found between TiO $T_{\rm eff}$'s and those from the other two methods are due to the behaviour of the TiO bands. The SpT sequence of late-types stars is classically defined by these bands. Thus, their strength had always been interpreted as a measurement of $T_{\rm eff}$. \cite{dav2013} concluded that an analisys based on TiO bands does not allow a correct derivation of $T_{\rm eff}$, because these bands do not only respond to temperature, but are also affected by 3D effects and luminosity. The $T_{\rm eff}$'s derived by \cite{dav2013} through their preferred SED method span a surprisingly narrow range ($4\,150\pm150\:$K), regardless of the parental galaxy. From this, they suggested that the typical $T_{\rm eff}$ of RSGs of any given population does not depend on the metal content. They also suggested, contrary to the classical interpretation, that the SpT of RSGs is only dependent on their evolutionary state. Thus, they imply that all RSGs have roughly the same $T_{\rm eff}$, regardless of their SpT.

Since \cite{dav2013} published their results, many other works have studied small samples of RSGs in different environments following the method originally proposed by \cite{dav2010}: atmospheric parameters are derived from the fit of a few atomic features in the $J$-band, where RSGs do not present strong molecular absorptions. This method has the advantage that it only uses photospheric features instead of molecular bands that are produced higher in the atmosphere. Therefore, their results are not affected by strong molecular absorptions. All these works found similar $T_{\rm eff}$ ranges for their samples, regardless of the typical metallicity of the parental environment: from $3\,800\:$K to $4\,100\:$K for the sample from Perseus OB1, in the Milky Way \citep{gaz2014}; from $3\,790\:$K to $4\,000\:$K for NGC\,6882 \citep{pat2015}; from $3\,760\:$K to $4\,030\:$K for NGC\,2100, in the LMC \citep{pat2016}; and from $3\,800\:$K to $4\,200\:$K in the re-analysis of the Magellanic sample presented in \cite{dav2013} with this method, which does not find any difference between the samples from the two galaxies \citep{dav2015}. Finally, \cite{gaz2015} studied RSGs from the spiral galaxy NGC\,300 at multiple radial distances, and did not find any trend between metallicity (which covers $0.6$~dex) and typical $T_{\rm eff}$. All these works agree with the conclusion of \cite{dav2013}, because of the similarity of the $T_{\rm eff}$ ranges found.

Last year, we presented our analysis of two very large samples of cool supergiants (CSGs), one from the LMC and the other from the SMC \citep{dor16a}. In total, our sample had more than 500~CSGs, including classical RSGs (with K and M types) but also G~supergiants, which we concluded are a significant part of the same population of evolved high-mass stars as RSGs in the SMC. Our main objective was to test the hypothesis of \cite{dav2013} about the lack of dependence between SpT and $T_{\rm eff}$, and the relation of SpT with stellar luminosity and evolutionary stage (usually related to mass loss). For that work, we applied the recommendations of \cite{dav2013} about the necessity of using atomic features from the photosphere for the spectral analysis instead of TiO bands. Therefore we used a long list of well-known atomic lines in the range of the infrared Calcium Triplet (CaT), from $8\,400\:$\AA{} to $8\,900\:$\AA{}. However, we did not derive stellar parameters directly from these lines. Instead, we studied their global behaviour, by comparing it with the theoretical dependence of line strengths on $T_{\rm eff}$, luminosity, and metallicity. Our results provide indirect proof, although with high statistical significance, that there is a $T_{\rm eff}$ scale in CSGs; in other words, that SpT depends mainly on $T_{\rm eff}$. However, we also found a secondary but clear dependence on luminosity (bolometric magnitude), which confirms for a significantly large sample a long suspected effect: more luminous CSGs typically tend to present later SpTs. 

In this work we go a step further than previous works. We calculate $T_{\rm eff}$ and [M/H] for most of the RSGs in the sample of \citet{dor16a}, which is much larger than any other sample previously used for this purpose. For this, we performed a careful empirical selection of atomic features. Our main objective is to prove if a connection between $T_{\rm eff}$ and SpTs exists. We take advantage of the statistically significant size of our sample. In addition, we want to study the relation between the different stellar parameters in the CSG populations of both MCs. Finally, we will compare our results with theoretical evolutionary tracks of massive stars.

The sample that we employ in the present work is described in Sect.~\ref{sample}. Detailed explanations of the derivation of stellar parameters are given in Sect.~\ref{spec_analys}. In Sect.~\ref{discus} we discuss our results and compare them to those obtained in previous works, as well as to evolutionary tracks. Finally in Sect.~\ref{conclus}, we summarize our conclusions about the existence of an $T_{\rm eff}$ scale for the CSGs studied in the present manuscript.

\section{The sample}
\label{sample}

The sample of CSGs analysed in this work has already been published in \citet[Paper~I]{gon15}, \citet[Paper~II]{dor16a} and \citet[Paper~III]{dor16b}. In \citetalias{gon15} we presented and discussed the selection of targets, the observations and our classification using optical spectra. In \citetalias{dor16a}, we analysed the sample of CSGs. We studied the empirical behaviour of atomic features present in the infrared Calcium Triplet (CaT) spectral region against SpT, luminosity, and mass loss. We also discussed the spectral variability of stars in our sample, their SpT distributions, and the implications of our results for the current understanding of CSGs. In \citetalias{dor16b}, we employed these spectral features mentioned above to implement an automated method to identify CSGs through their spectra.

The observations are explained in detail in \citetalias{gon15}. Thus, here we will only summarize them. We used the fiber-fed dual-beam AAOmega spectrograph on the 3.9~m Anglo-Australian Telescope on four separate runs between 2010 and 2103. Thanks to the dual-beam, all the targets were observed simultaneously in the optical and the CaT spectral ranges. The grating used for the infrared range was 1700D, which provides a $500\:$\AA{} wide range, centred on $8\,600\:$\AA{} in 2010 and on $8\,700\:$\AA{} for all other epochs. This grating has a nominal resolving power ($\lambda/\delta\lambda$) of $10\,000$ at the wavelengths considered. The gratings for the optical range have no bearing on the present work; their details are provided in table~1 of \citetalias{gon15}.

Spectra in the optical range were used to infer the SpT and luminosity class (LC) for all the targets observed \citepalias[see][for details]{gon15}. We used classical criteria based on atomic line ratios and TiO band depths, when present in the spectra. We also used radial velocities (RV) alongside our LC classification to confirm their membership to the MCs \citepalias[see figs.~5 and~6 from][]{gon15}.  Some targets were observed at least twice within the same epoch. If so, final SpT and LC classifications for each target at each epoch were calculated using a S/N-weighted average. 

Since the optical range is crowded with molecular features (mainly from TiO), the characterization of the spectra of these stars becomes difficult, due to effects like the lack of a visible continuum. We decided to tackle the problem in two fronts. Firstly, we restricted the spectral range to the CaT spectral region that is much less affected by TiO bands than the blue or red regions and contains a large number of well-resolved weak atomic lines. Secondly, we imposed a cut based on our previously derived SpTs. In the CaT range, molecular bands are not noticeable for types earlier than M1. Moreover, atomic lines become significantly  eroded by these absorption bands on the continuum only for types later than M3 \citepalias[see sect.~2.2 and fig.~1 in][]{gon15}. Therefore, here we only use stars with SpT M3 or earlier. This limitation excludes only 8.3\% of the CSGs (most of them from the LMC). The statistical significance of the sample is not compromised. While most previous works on this topic centre their efforts on K and M~supergiants, denominated as RSGs (See Sect.~\ref{intro}), we also include the G~supergiants in the sample analysed in this paper, since in \citetalias{dor16a} we demonstrated that these yellow supergiants (YSGs) are part of the same population as RSGs in the SMC. Thus, we refer to the stars in our sample as CSGs, which includes G, K and M types.
In addition, we truncated the sample by leaving out every spectrum with an average S/N lower than 30, in order to obtain a minimum quality for our derived stellar atmospheric parameters. In Section~\ref{fin_err_bud} we discuss further the effect of low S/N on the parameter uncertainties.

The bolometric magnitudes of our targets were calculated through the bolometric correction of \cite{bes1984}, which is derived from the $(J-K)$ color index. We used the 2MASS photometry \citep{skr2006} for this calculation, previously transforming it to the AAO system used by \cite{bes1984}. We have assumed that the effect of the reddening on this calibration is negligible \citepalias[see section~2.2 of][for more details]{dor16a}. The absolute bolometric magnitudes were calculated using the following distance modulus to the MCs: $\mu=18.48\pm0.05\:$mag for the LMC \citep{wal2012} and $\mu=18.99\pm0.07\:$mag for the SMC \citep{gra2014}.

\section{Spectroscopic analysis}
\label{spec_analys}

All spectra were normalized using the {\sc continuum} task within {\sc IRAF\footnote{IRAF is distributed by the National Optical Astronomy Observatories,which are operated by the Association of Universities for Research in Astronomy, Inc., under cooperative agreement with the National Science Foundation.}}. Later, we shifted the spectra into the rest-frame using the radial velocities already calculated in \citetalias{gon15}, through the {\sc IRAF dopcor} task.

\subsection{Stellar Parameters}
\label{sparam}
Stellar atmospheric parameters and abundances were computed using a previously generated grid of synthetic spectra (see Sect.~\ref{syn} and Table~\ref{parGTAB}). We employed a modified version of the automated code {\scshape StePar} \citep[see][]{tab12}. The new version relies upon spectral synthesis instead of equivalent widths (EWs). We also replaced the original optimization method (based on a downhill simplex algorithm) with a Metropolis-Hastings algorithm using Markov chains \citep{met53}. Our method generates a Markov-Chain of 20\,000 points starting from an arbitrary point. To be able to evaluate any point within the stellar parameter space, we employed a bilinear interpolation scheme. As objective function we used a $\chi$-squared in order to fit any previously selected spectral features.  

The present version of {\scshape StePar} allows the simultaneous derivation of any set of stellar atmospheric parameters. In this case we restricted them to only two variables, $T_{\rm eff}$ and metallicity ([M/H]). Surface gravity ($\log\,g$) was kept fixed to $0.0\:$dex, microturbulence ($\xi$) was also kept constant at $3\:$km\,s$^{-1}$, {according to the approximation discussed by \citet{gher07}.} Our analysis revolves around a few atomic features  in the spectral range around the CaT, from $8\,400$ to $8\,900\:$\AA{}. We employed some empirically selected lines of Mg, Si, Ti, and Fe (see Section~\ref{syn}). We also convolved our grid of synthetic spectra with a gaussian kernel (FWHM$\:\approx30\:$km\,s$^{-1}$) to account for the instrumental broadening.

\subsection{Synthetic spectra}
\label{syn}

The synthetic spectra were generated using two sets of one-dimensional LTE atmospheric models, namely: ATLAS-APOGEE (KURUCZ) plane-parallel  models \citep{mes12}, and MARCS spherical models with $15\:$M$_{\odot}$ \citep{gus08}. The radiative transfer code employed was \textit{spectrum} \citep{graco94}. Also, we employed the abundances of \citet{asp05} as Solar reference (namely, $Z_{\odot}$~$=$~0.0122). Although MARCS atmospheric models are spherical, \textit{spectrum} treats them as if they were plane-parallel. This can lead to a small inconsistency in the synthetic spectrum calculations. However, the study by \citet{hei06} concluded that any difference introduced by the spherical models in a plane-parallel transport scheme is not significant.  As line list, we employed a selection from the VALD database \citep{pis95,kup00,rya15}, taking into account all the relevant atomic and molecular features (dominated by TiO and CN) that can appear in CSGs with SpT M3 or earlier. Atomic data for the most prominent atomic features and synthesis ranges employed can be found in Table~\ref{linTab}. In addition, as Van der Waals damping prescription we employed the Anstee, Barklem, and O'Mara theory (ABO), when available in VALD \citep[see][]{bar00}. The grid of synthetic spectra was generated for a single $\log{g}=0\:$dex. Effective temperature $T_{\rm eff}$ ranges from $3\,500\:$K to $6\,000\:$K with a step of $250\:$K for the spectra generated using KURUCZ atmospheric models, whereas, for MARCS synthetic models, $T_{\rm eff}$ varies from $3\,300\:$K to $4\,500\:$K; the step is $250\:$K above $4\,000\:$K and $100\:$K otherwise. The microturbulence ($\xi$) was fixed to $3\:$km\,s$^{-1}$, see Sect.~\ref{sparam}. Finally, the metalicity ranges from [M/H]$=-1.5\:$dex to [M/H]$=1.0\:$dex in $0.25\:$dex steps for KURUCZ models, whereas MARCS models cover only from [M/H]$=-1.0\:$dex to [M/H]$=0.5\:$dex, in $0.25\:$dex increments.

\subsection{Final results and error budget}
\label{fin_err_bud}

\begin{figure}
	\centering
	\includegraphics[trim=1cm 0.5cm 1.8cm 1.2cm,clip,width=8.5cm]{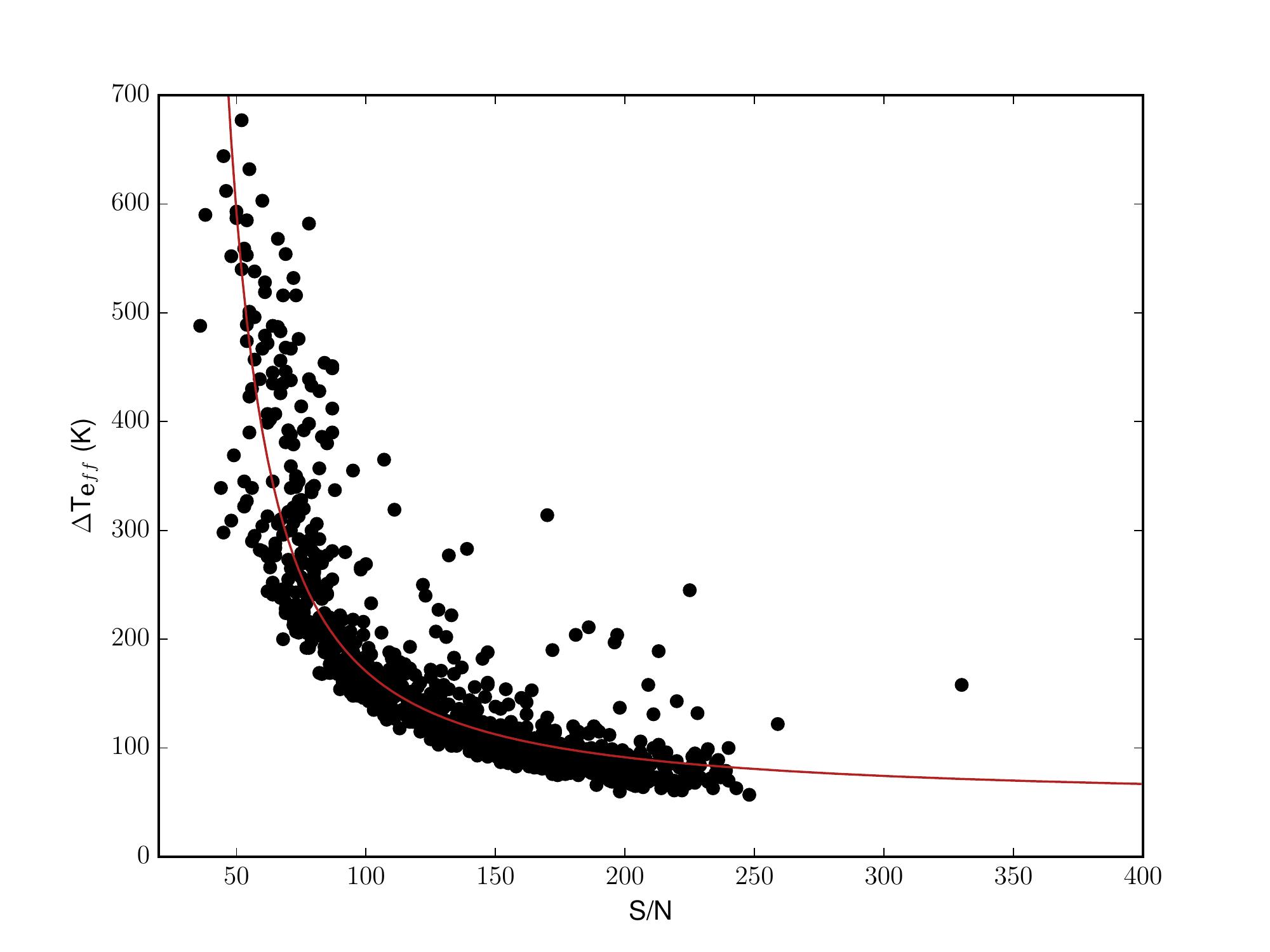}
	\includegraphics[trim=1cm 0.5cm 1.8cm 1.2cm,clip,width=8.5cm]{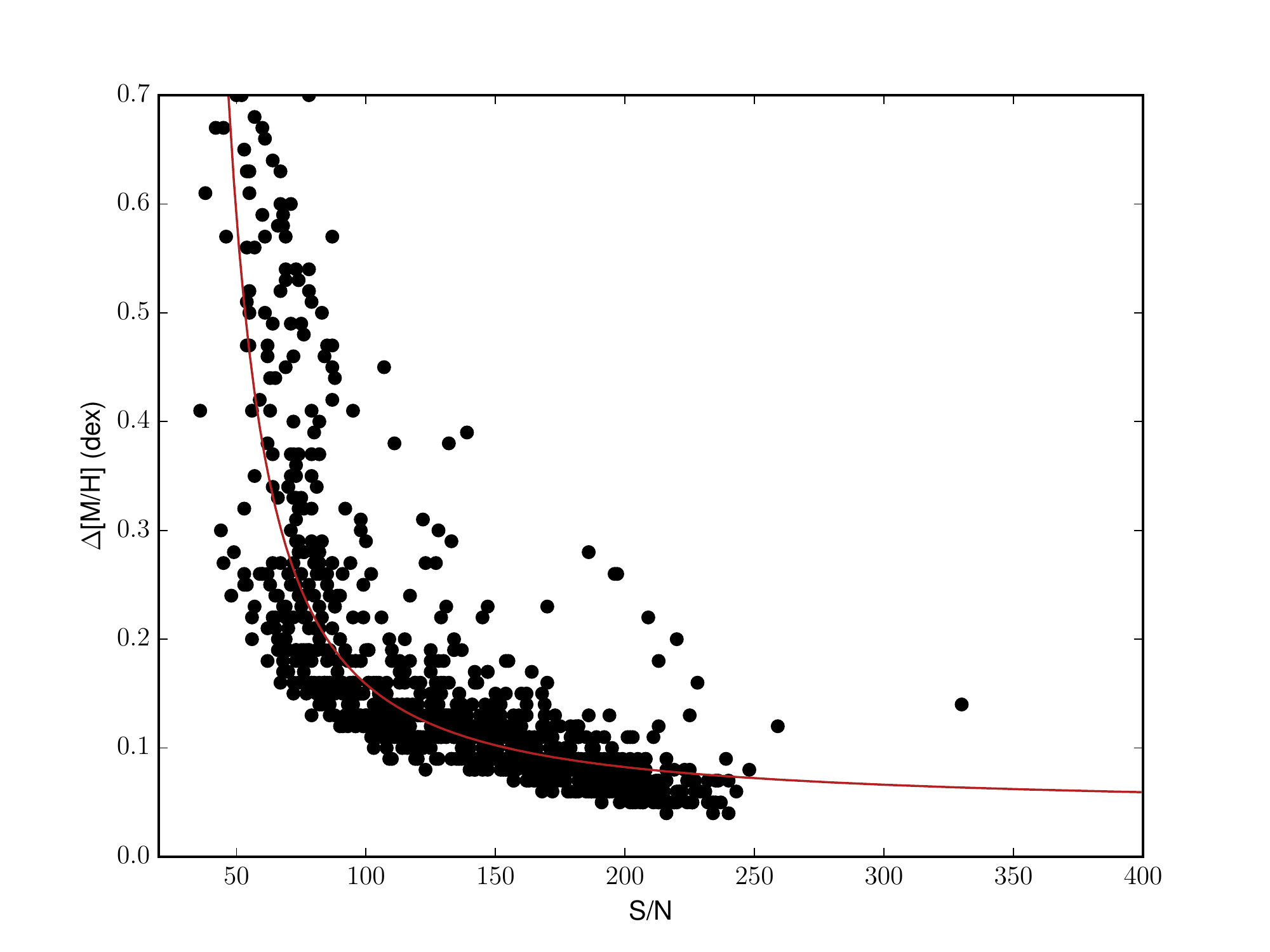}
	        \caption{Derived $\Delta$T$_{\rm eff}$ and $\Delta$[M/H] against S/N. The data plotted are the results obtained for all the individual spectra in our sample, without taking into account that some stars have multiple spectra. The red line represents the fit to the behaviour of uncertainties as described in Sect.~\ref{fin_err_bud}.}
		\label{errpar}
\end{figure}

Uncertainties on the stellar parameters are derived from the statistics of the Markov Chain. Firstly, we chose only the second half of the chain. Secondly, we divided the chain into sequences of five points, and we took only the first point of each sequence. Then we generated histograms with those points using either $T_{\rm eff}$ or [M/H]. Errors and mean values were obtained by fitting a gaussian to each histogram. Points were combined into intervals according to the rule described in \citet{fdr81}. Typical errors are shown in Table~\ref{errTab}. In addition, we show our best-fits synthetic models for seven high S/N spectra in Fig.~\ref{fitsyn}.
In general terms, our internal uncertainties are dominated by two main sources. The first contribution comes from the assumptions intrinsic to any stellar parameter determination (atmospheric models, lines, and broadening, to name a few). The second one is, roughly speaking, data quality. The latter is dominated by S/N (see Table~\ref{errTab}), which can make uncertainties escalate to higher values as S/N decreases. However, if S/N is high enough, any calculation is driven by the methodology employed to derive stellar parameters \citep{rec06}. In Fig.~\ref{errpar}, we plot uncertainties on $T_{\rm eff}$ and [M/H] as a function of S/N, for all the spectra analysed. We assumed that our uncertainties behave as $\alpha e^{\frac{\beta}{\rm S/N}}$. Using least squares, we can calculate $\alpha$, which corresponds to the minimum uncertainty. We obtain $49\:$K and $0.04\:$dex as lower limits to this method (assuming that S/N tends towards $\infty$).

\begin{table}
	\centering
	\caption{Typical uncertainties of the sample.}
	\label{errTab}
	\begin{tabular}{lcccc}
		\hline
		S/N & $<50$ & $\geq50$ & $\geq100$ & All \\
		\hline
		$\Delta T_{\rm eff}$ (K)  & 493 & 169 & 114 & 172 \\
		$\Delta$[M/H] (dex) & 0.47 & 0.16 & 0.11 & 0.17 \\
		Spectra & 10  & 905  & 629  & 915 \\
		\hline
	\end{tabular}
\end{table}

In addition, our internal errors can tell us if a result is well defined or not. This is specially important for points close to the grid limits. We imposed the condition that a point is good only if its 1-$\sigma$ levels fall within the limits of our grids. This removes spurious results close to the grid edges, which can lead to low-precision stellar atmospheric parameters.

\begin{figure}
\centering
\includegraphics[trim=1cm 0.4cm 1.8cm 1.2cm,clip,width=8.5cm]{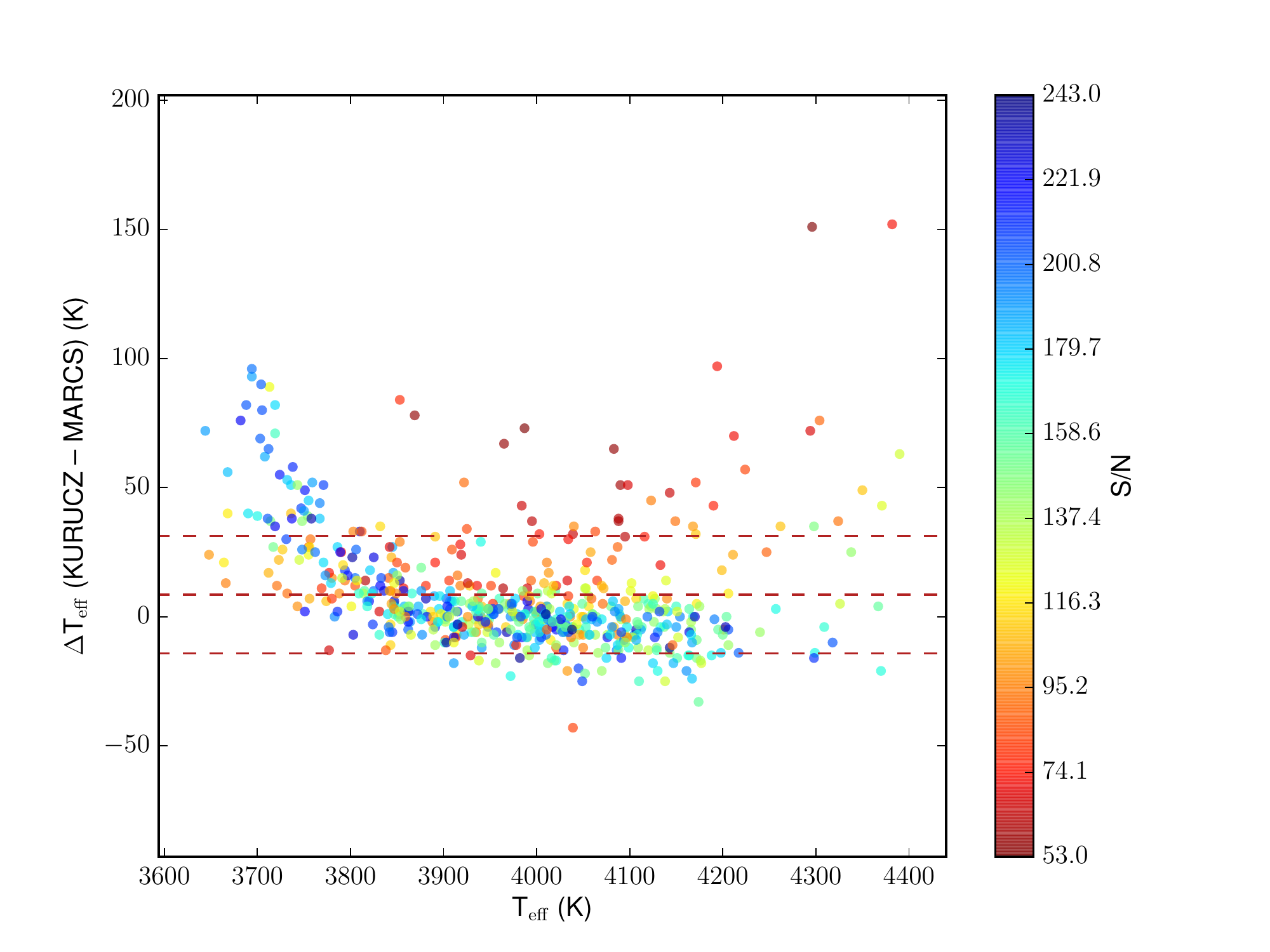}
\includegraphics[trim=1cm 0.4cm 1.8cm 1.2cm,clip,width=8.5cm]{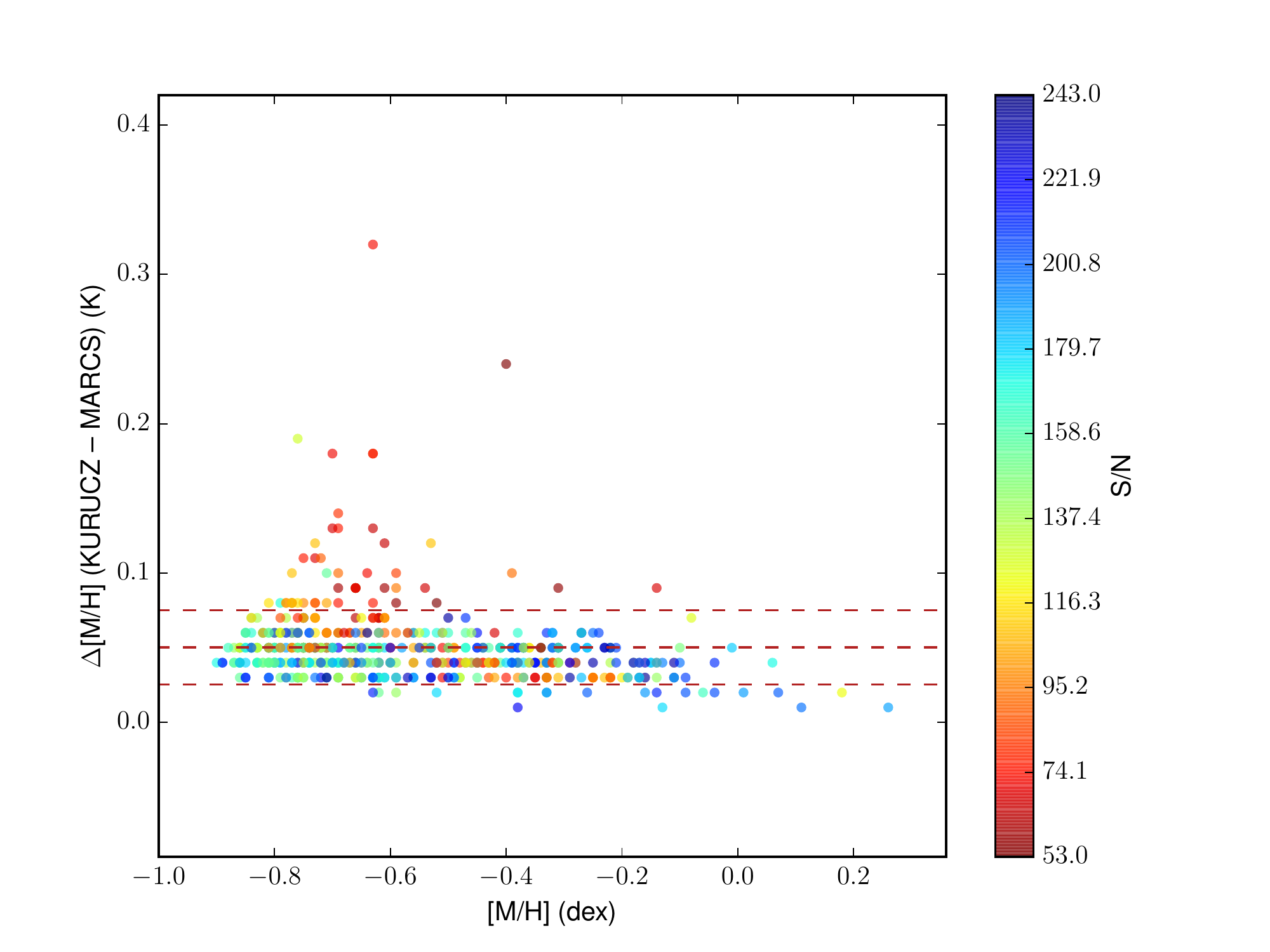}
\caption{ $T_{\rm eff}$ and [M/H] differences when calculated using either KURUCZ or MARCS atmospheric models against combined values. All data plotted are the results obtained for all the individual spectra in our sample, even if they belong to the same star in the same epoch. Red dashed lines depict the average difference value and the 1-$\sigma$ levels.}
\label{kvsm}
\end{figure}

Another important point is how to combine calculations coming from two different atmospheric model grids. We do not find a significant systematic offset between KURUCZ and MARCS models above $4\,000\:$K. However, as we clearly show in Fig.~\ref{kvsm}, an offset is noticeable for $T_{\rm eff}$ below $4\,000\:$K. This offset must be taken with care, since at low $T_{\rm eff}$, the differences raise linearly and up to $100\:$K. This shows an important issue regarding on how stellar atmospheric models were calculated. MARCS atmospheric models for $15\:$M$_{\odot}$ were computed using spherical geometry. In spite of that fact, if we were to average those differences they will be at most $50\:$K in the worst case scenario\footnote{We are in no position to state which atmospheric models are better. Of course, it can be argued that MARCS models are spherical; KURUCZ models, however, cover a wider atmospheric extension. Each approximation has its own assumptions, and unfortunately we have to deal with them in a reasonable manner.}. Therefore, our internal uncertainties are compatible with that systematic effect. Global differences show a small average value of $8\pm22\:$K. Thus, our final results are an average of KURUCZ and MARCS calculations, whenever both grids produce a well defined set of stellar atmospheric parameters for a given individual spectrum. For [M/H] we obtained a small difference of $0.05\pm0.02\:$dex, indicating that [M/H] is very stable regardless of the model grid employed for the calculations.

\begin{figure}
\centering
\includegraphics[trim=1cm 0.4cm 1.6cm 1.2cm,clip,width=8.5cm]{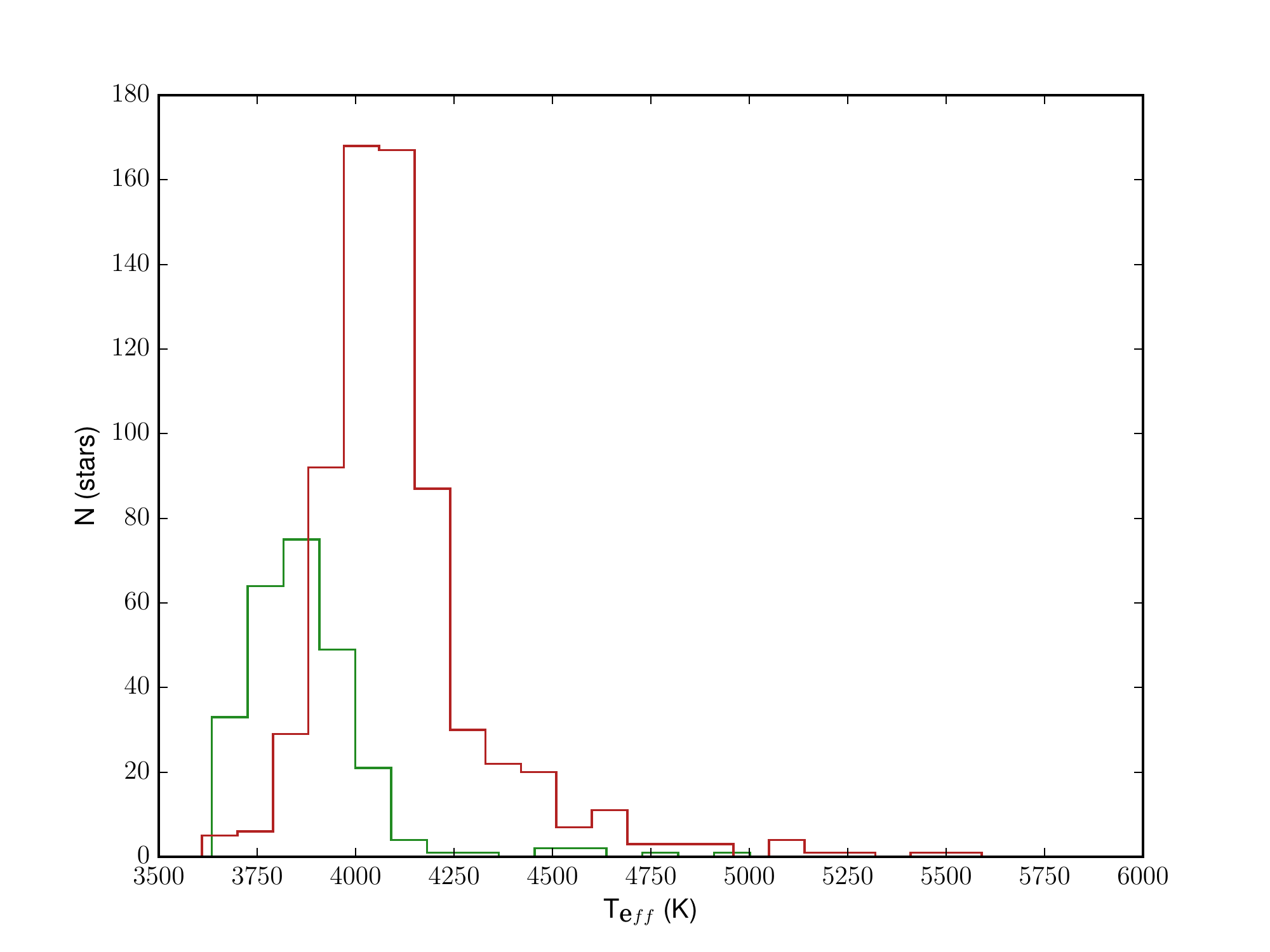}
\includegraphics[trim=1cm 0.4cm 1.6cm 1.2cm,clip,width=8.5cm]{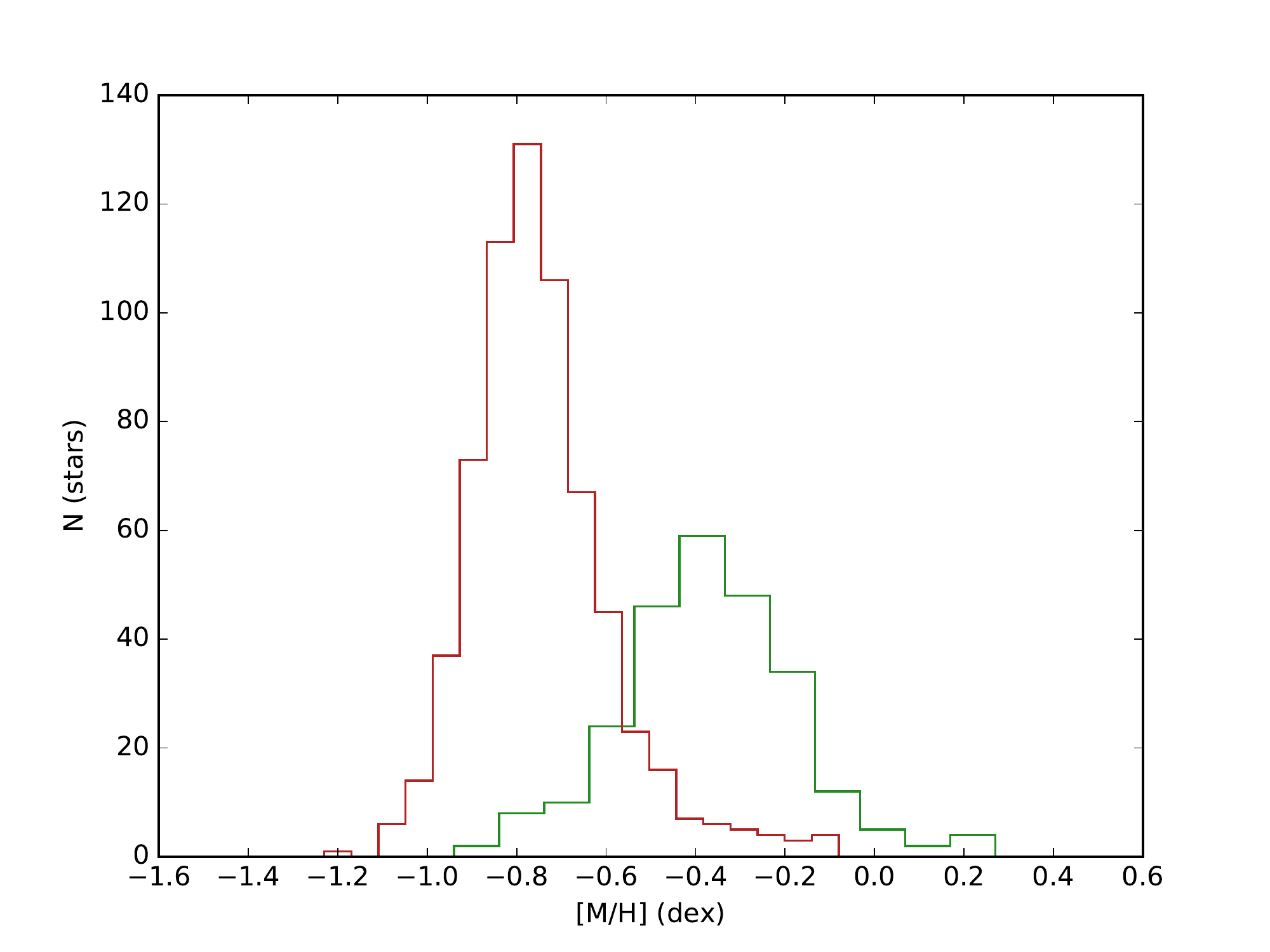}
	\caption{Derived $T_{\rm eff}$ and [M/H] for the whole sample. LMC stars are plotted in red, whereas SMC stars are represented by a green line. The bin size was determined using the \citet{fdr81} rule. The data obtained for spectra of the same star in different epochs are not combined, but treated as different stars for these figures.}
\label{histpar}
\end{figure}

There are stars in our sample with more than one spectrum taken in the same epoch (i.e.\ within two nights of each other). We calculated the parameters separately for each spectrum available. Then, we combined the results for spectra of the same star in the same epoch, calculating their average $T_{\rm eff}$ and [M/H] weighted by the S/N of the spectra. The histograms of the results are shown in Fig.~\ref{histpar}. We did not combine spectra from different epochs of the same star in these figures because a large number of CSGs present spectral variability \citepalias[see][for details]{dor16a}. We also obtained the internal uncertainties for both temperature and metallicity. For this we used all the spectra of stars which do not show spectral variability. In the case of variable targets, as the variations typically have periods of hundreds of days, we compared spectra only inside each epoch. We obtained an internal dispersion of $25\:$~K for $T_{\rm eff}$, and $0.06\:$~dex for [M/H]. These values are well below our internal uncertainties, thus proving that our method is stable when deriving stellar parameters of the same object in different epochs as long as it is  not variable in SpT. We discuss the effects caused by SpT variability in detail in Sect.~\ref{spec_var}.

In order to evaluate  the effect of fixing the value of microturbulence, we calculated two other sets of stellar parameters by varying microturbulence by $\pm$~0.5~km~s$^{-1}$ (see Figs.~\ref{xiKUR} and~\ref{xiMAR}). Metallicities only vary by a factor of approximately $\mp$~0.2 dex for the MARCS grid and $\mp$~0.18~dex for the KURUCZ grid. However, the effective temperature was stretched by a small percentage:  $\mp$~15~$\%$ for MARCS, and $\mp$~5~$\%$  for KURUCZ. The variations found do not alter any tentative temperature scale, since any correlation that we may want to calculate is scale-invariant.

Our derived average [M/H], weighted by the S/N of each individual spectrum in our sample (i.e.\ including duplicated observations), is $-0.35\pm0.15\:$dex for the LMC, and $-0.75\pm0.11\:$dex for the SMC. \citet{dav2015} made a careful and detailed compilation of literature values for each MC. They reported an average present-day metallicity range from $-0.2$ to $-0.4$ dex for the LMC, whereas the SMC ranges from $-0.5\:$dex to $-0.8\:$dex. These values are representative of what can be found in the literature \citep[see][and references therein]{dav2015}. Thus, our results are fully consistent with what previous studies have found for each cloud, taking into account that \citet{dav2015} mention that these typical values of [M/H] for each MC are dependent on the sample included in each study.

\subsection{LTE systematics}
\label{lte}

Our calculations are purely based on an LTE classical but conservative approach. Some recent works have made use of NLTE corrections to LTE calculations \citep[e.g.][]{dav2015,gaz2015,pat2016}. To estimate the validity of our results, we need to know the systematics of our LTE calculations against NLTE methods. NLTE corrections for RSGs have been quantified in the literature for some chemical elements: \ion{Fe}{i}, \ion{Ti}{i} \citep{ber12}, \ion{Si}{i} \citep{ber13}, and \ion{Mg}{i} \citep{ber15}.

\begin{figure*}
\centering
\includegraphics[trim=0.5cm 0.4cm 1.6cm 1.2cm,clip,width=8.5cm]{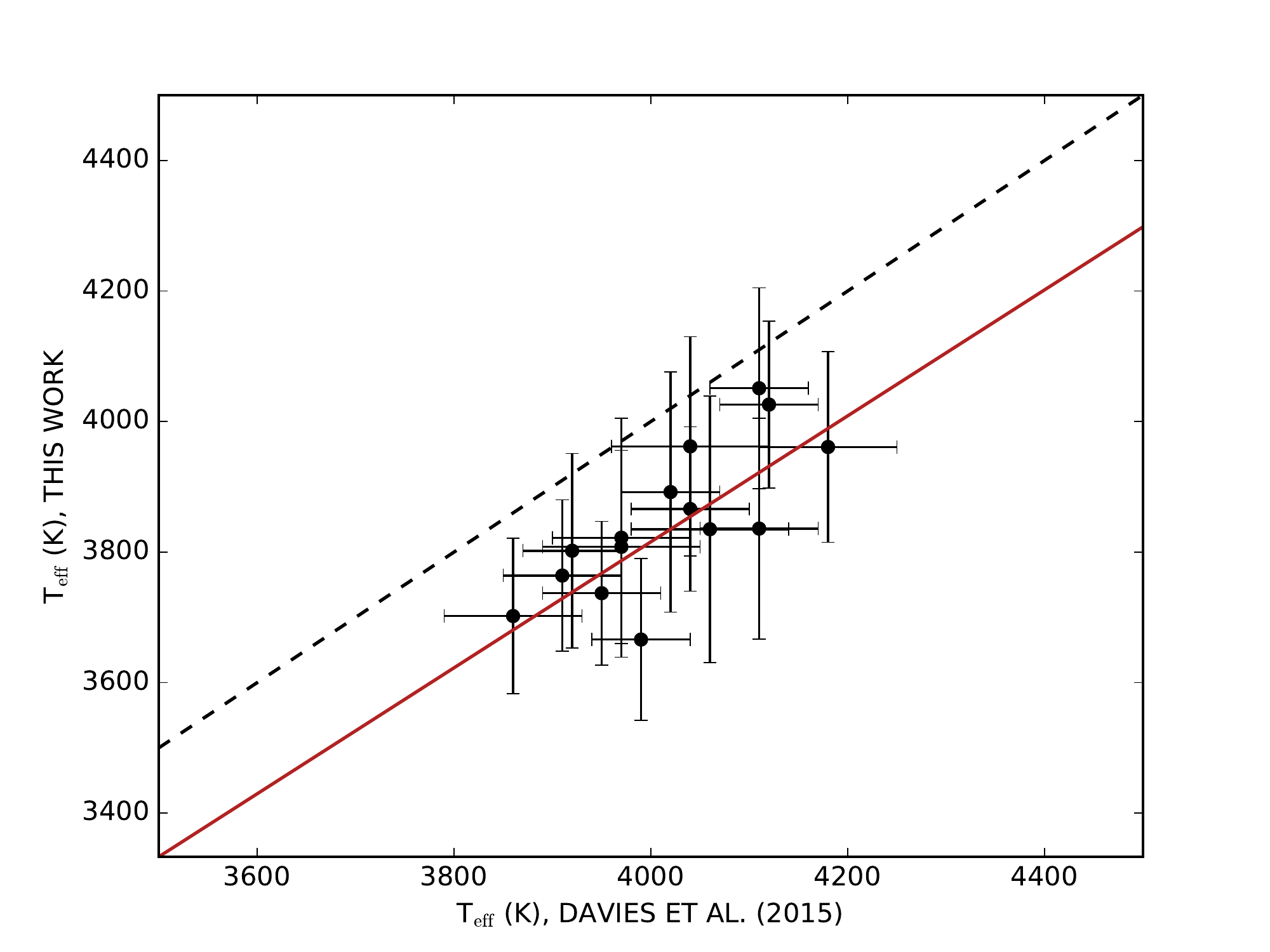}
\includegraphics[trim=0.5cm 0.4cm 1.6cm 1.2cm,clip,width=8.5cm]{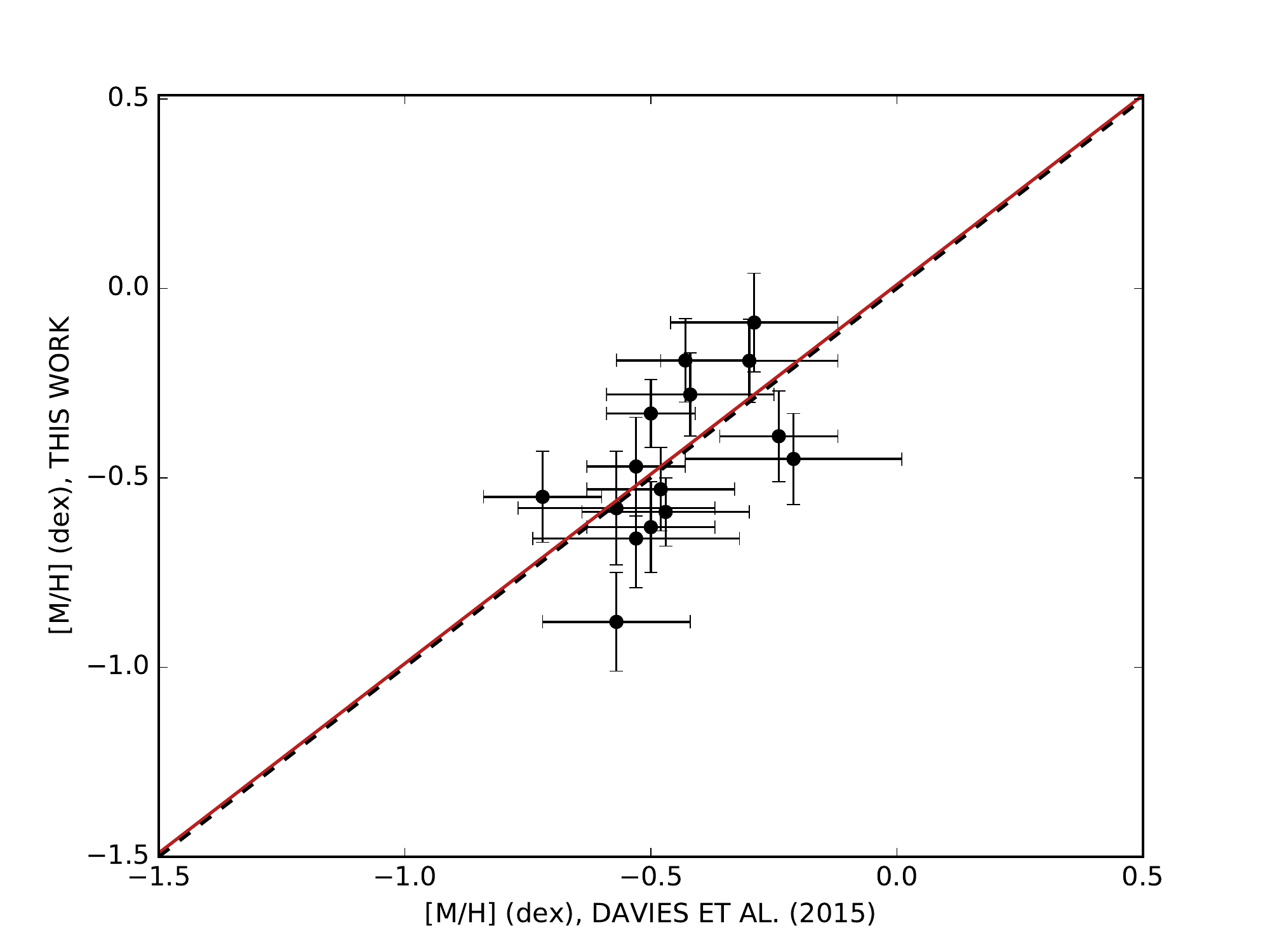}
        \caption{Our stellar atmospheric parameters vs those analysed by \citet{dav2015} . The dashed line represents a 1:1 relation, whereas the solid line is simply a linear fit trough the points.}
\label{davgraf}
\end{figure*}

\begin{table*}
        \centering
	\caption{Our calculations using the X-shooter spectra analysed by \citet{dav2013,dav2015}. The names and stellar parameters of these stars are those used in \citet{dav2015} (Dav15), but they are referred to the catalogue of \citet{mas2002}. The {\scshape StePar} parameters are those obtained in the present work.}
        \label{nltevslte}
               \begin{tabular}{lccccc}
        \hline
        NAME  & Galaxy & $T_{\rm eff}$~({\scshape StePar})  & [M/H]~({\scshape StePar}) & $T_{\rm eff}$~(Dav15)  & [M/H]~(Dav15) \\
        \hline
       064048 & LMC & $3702\pm119$ & $-0.28\pm0.11$ & $3860\pm70$ & $-0.42\pm0.17$ \\
       067982 & LMC & $3764\pm116$ & $-0.19\pm0.11$ & $3910\pm60$ & $-0.43\pm0.14$ \\
       116895 & LMC & $3737\pm110$ & $-0.19\pm0.11$ & $3950\pm60$ & $-0.30\pm0.18$ \\
       131735 & LMC & $4051\pm154$ & $-0.33\pm0.09$ & $4110\pm50$ & $-0.50\pm0.09$ \\
       137818 & LMC & $3666\pm124$ & $-0.63\pm0.12$ & $3990\pm50$ & $-0.50\pm0.13$ \\
       143877 & LMC & $3835\pm204$ & $-0.09\pm0.13$ & $4060\pm80$ & $-0.29\pm0.17$ \\
       011709 & SMC & $3892\pm184$ & $-0.47\pm0.13$ & $4020\pm50$ & $-0.53\pm0.10$ \\
       020133 & SMC & $3808\pm148$ & $-0.45\pm0.12$ & $3970\pm80$ & $-0.21\pm0.22$ \\
       021362 & SMC & $3822\pm183$ & $-0.58\pm0.15$ & $3970\pm70$ & $-0.57\pm0.20$ \\
       030616 & SMC & $3866\pm126$ & $-0.59\pm0.09$ & $4040\pm60$ & $-0.47\pm0.17$ \\
       034158 & SMC & $3961\pm146$ & $-0.53\pm0.11$ & $4180\pm70$ & $-0.48\pm0.15$ \\
       035445 & SMC & $3962\pm168$ & $-0.66\pm0.13$ & $4040\pm80$ & $-0.53\pm0.21$ \\
       049478 & SMC & $3836\pm169$ & $-0.39\pm0.12$ & $4110\pm60$ & $-0.24\pm0.12$ \\
       050840 & SMC & $3802\pm149$ & $-0.55\pm0.12$ & $3920\pm50$ & $-0.72\pm0.12$ \\
       057386 & SMC & $4026\pm128$ & $-0.88\pm0.13$ & $4120\pm50$ & $-0.57\pm0.15$ \\
        \hline
\end{tabular}
\end{table*}

To accomplish this particular aim, we downloaded the spectra analysed by \citet{dav2013,dav2015},  which are available under ESO program 088.B-0014(A). As there were multiple spectra for each star, we chose only those with $R\approx7\,000$, of which there is only one per star. We chose this resolution because it is closest to that of our own sample. For consistency, we did not use those stars that had not been observed at this resolution. Then we applied our analysis to the same region of the $I$-band spectrum that we have used for our sample, as explained in previous sections. Before this, we also calculated our own RV for the stars, using the CaT spectral range, in the same way as for the sample of this work \citepalias[see][]{gon15}. In Table~\ref{nltevslte} and Fig.~\ref{davgraf} we show a comparison between both methodologies. We obtain an average difference in $T_{\rm eff}=-168\pm177\:$K. For [M/H] the offset is $0.00\pm0.17\:$dex. Thus, our analysis results in lower effective temperatures and similar metallicities, see Fig.~\ref{davgraf}. However, this comparison must be taken with care. Even though we assume that NLTE is a better approximation than LTE, the differences found are compatible with zero if we simply consider the standard deviation. To evaluate its statistical significance, we employed a BIC analysis (bayesian information criterion) to the $T_{\rm eff}$ difference. The result of this analysis is $\Delta$BIC$=$BIC(offset)$-$BIC(no offset)$=1.77$. Thus, since $\Delta$BIC is less than 2, we cannot ascertain that the difference between the NLTE and LTE temperatures is statistically significant, at least for this small comparison sample.

Our line diagnostics are dominated by \ion{Ti}{i} and \ion{Fe}{i}. \citet{ber12} reported that NLTE corrections for \ion{Ti}{i} can be really important (up to $0.4\:$dex), whereas, NLTE \ion{Fe}{i} corrections are smaller, reaching at most $0.15\:$dex. \ion{Si}{i} NLTE corrections \citep{ber13} are at a level similar to those of \ion{Fe}{i}. \ion{Mg}{i} also presents similar effects as \ion{Fe}{i} and \ion{Si}{i} \citep{ber15}. All these works \citep{ber12,ber13,ber15} report that global [M/H] calculations (as those presented here) can be affected at the level of $\Delta$~[M/H]~$\approx0.15\:$dex. Additionally, they also study the effects of NLTE on line profiles. As one might expect, line cores are more affected. Line cores form in the upper levels of the atmosphere, where the LTE approximation starts to deviate from "reality", this effect can be seen in Fig.~\ref{fitsyn} for only two lines. Our best fits are off for some lines around 8800\AA~because the LTE approach does not hold for the upper layers of the stellar atmosphere. However, our internal uncertainties are sufficiently conservative to account for any possible differences (see Table~\ref{errTab}).

\section{Discussion}
\label{discus}

\subsection{The temperature scale of the Magellanic cool supergiants}
\label{teff_scale}

\subsubsection{Correlations between effective temperature and spectral type}
\label{corr_spt_teff}

The large size of our sample allows us to study the $T_{\rm eff}$ scale drawn by CSGs in the MCs with an unprecedented statistical significance. We used the sample described in \citetalias{dor16a}, which comprises data from 2012 (SMC) and 2013 (LMC). It does not include any observations from earlier epochs (2010 and 2011). Firstly, we cannot combine many spectra from different epochs, given the spectral variability we detected. Secondly, essentially all the stars from the 2010 and 2011 runs were also observed in 2012 and 2013 and thus we are not losing any stars, while preventing duplicities. In total, there are 527~CSGs observed in these two epochs. Unfortunately, for some stars we do not present any stellar parameters. We discarded stars later than M3, as well as those for which we do not find good stellar parameters, according to the criterion given in Sect.~\ref{fin_err_bud}. In total we characterized 445~CSGs, which represent $84$\% of the sample contained in \citetalias{dor16a} (see Table~\ref{parGTAB}). 

\begin{table*}
\caption{Pearson ($r$) and Spearman ($r_{\rm s}$) coefficients obtained for the correlations between $T_{\rm eff}$ and SpT in different subsamples. The values marked as Montecarlo are the average values and the standard deviations obtained from the $10\,000$ samples generated through Montecarlo (see text for details). Some subsamples were limited by spectral type or by $M_{\rm bol}$ (see text for a detailed explanation).}
\label{corr_table}
\centering
\begin{tabular}{c | c c | c c | c}
\hline\hline
\noalign{\smallskip}
&\multicolumn{2}{c|}{From Montecarlo}&\multicolumn{2}{c}{Original sample}&Size of\\
Sample&$r\pm\sigma_{\rm p}$&$r_{\rm s}\pm\sigma_{\rm s}$&$r$&$r_{\rm s}$&the sample\\
\noalign{\smallskip}
\hline
\noalign{\smallskip}
SMC&$-0.771\pm0.022$&$-0.76\pm0.03$&$-0.771$&$-0.82$&257\\
LMC&$-0.888\pm0.013$&$-0.855\pm0.021$&$-0.888$&$-0.811$&188\\
LMC (K--M types)&$-0.789\pm0.024$&$-0.78\pm0.03$&$-0.789$&$-0.79$&182\\
Both MCs&$-0.852\pm0.011$&$-0.854\pm0.013$&$-0.852$&$-0.914$&445\\
SMC ($M_{\rm bol}<-6\:$mag)&$-0.782\pm0.023$&$-0.77\pm0.03$&$-0.783$&$-0.81$&212\\
SMC ($M_{\rm bol}<-6.7\:$mag)&$-0.76\pm0.03$&$-0.74\pm0.04$&$-0.76$&$-0.79$&140\\
LMC ($M_{\rm bol}<-6\:$mag)&$-0.887\pm0.014$&$-0.850\pm0.023$&$-0.888$&$-0.803$&165\\
\noalign{\smallskip}
\hline
\noalign{\smallskip}
\end{tabular}
\end{table*}

Our sample covers a broad range in SpT (from G0 till M3) thus allowing us to ascertain the existence of a temperature scale. To this aim, we calculated the correlation coefficients of Pearson~($r$) and Spearman~($r_{\rm s}$) between SpT and $T_{\rm eff}$. While $r$ is sensible to linear correlations, $r_{\rm s}$ is more robust and can manage non-linear correlations, if any. We computed $r$ and $r_{\rm s}$ for each MC alone and for both galaxies together. To propagate uncertainties, we used a Montecarlo method. We created 10\,000 artificial samples of the same size as our own sample. Each artificial datapoint (i.e. an $T_{\rm eff}$ and SpT pair) in a generated sample was calculated by drawing it randomly from a two dimensional normal distribution centred on the values of one of the original measurements with a width given by the known uncertainties in each variable. Once these random samples were generated for all of our measurements, we calculated the correlation coefficients ($r$ and $r_{{\rm s}}$) for each one of the 10\,000 realizations, and analysed their distribution. The final values are the averages of the correlations coefficients found for the 10\,000 artificial samples and their uncertainties are given by their standard deviation. The results of this process can be seen in Table~\ref{corr_table}.

The values of $r$ and $r_{\rm s}$ obtained from our original sample indicate that SpT and $T_{\rm eff}$ are strongly correlated for CSGs, at least in the case of the MCs. The Montercalo-calculated $r$ and $r_{\rm s}$ remain similar to those of the original sample, in most cases both being inside the 2--$\sigma$ interval. These results confirm the existence of the correlation, even when our relatively high uncertainties in $T_{\rm eff}$ are taken into account.

\begin{figure*}
       \centering
       \includegraphics[trim=0.5cm 0.5cm 1.8cm 1.2cm,clip,width=8.8cm]{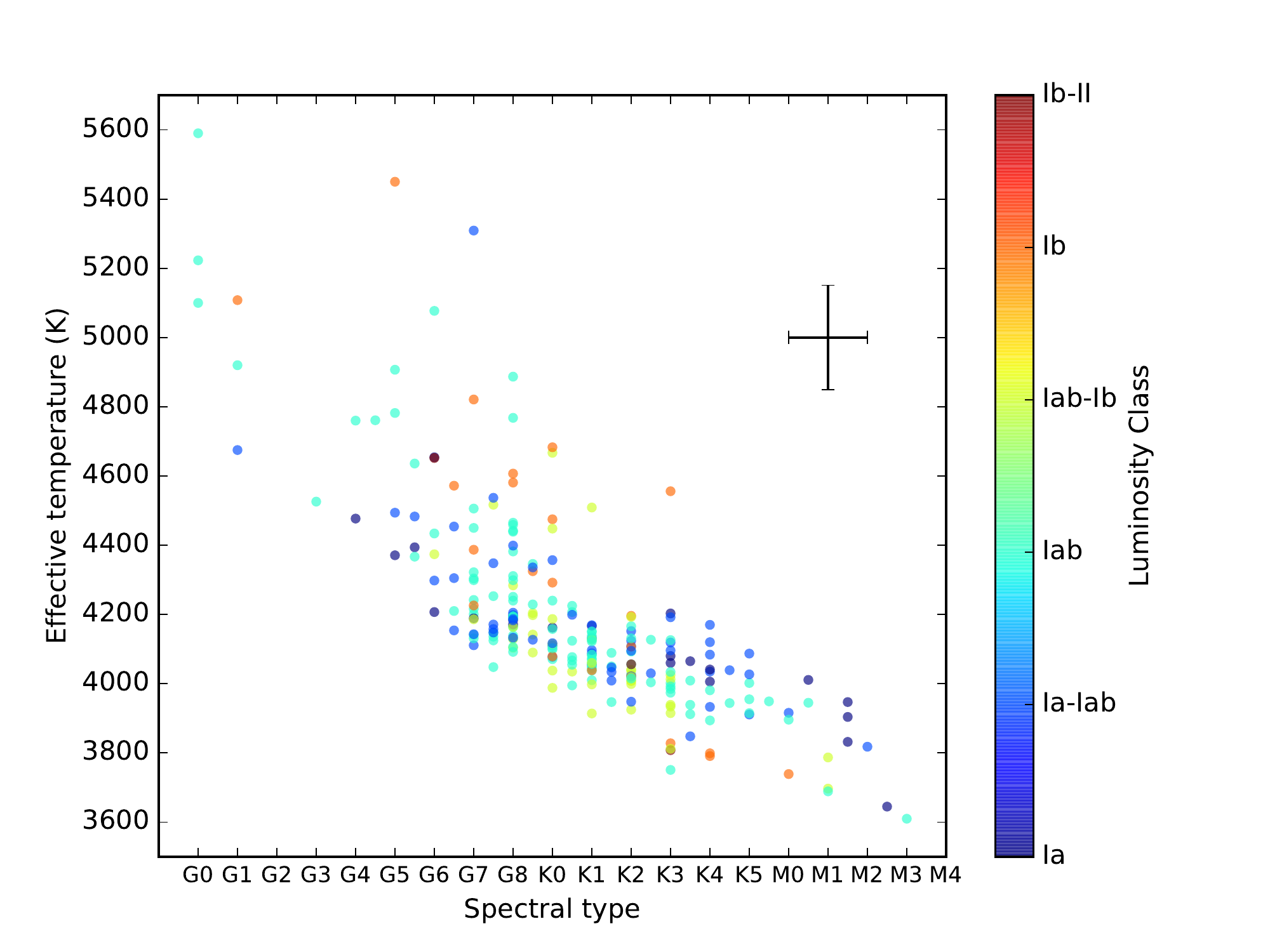}
       \includegraphics[trim=0.5cm 0.5cm 1.8cm 1.2cm,clip,width=8.8cm]{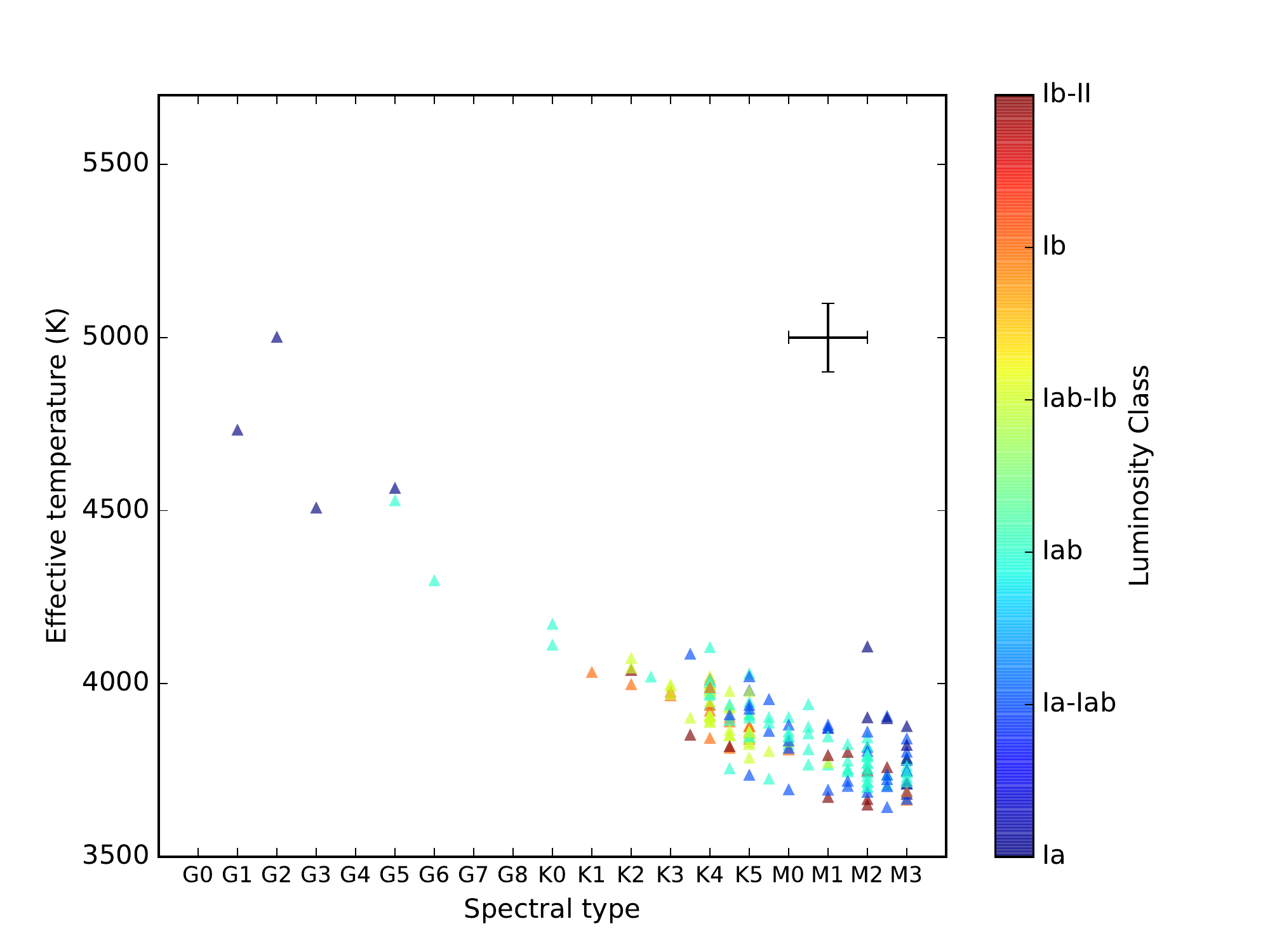} 
       \caption{Effective temperature against spectral type. The colour indicates luminosity class. The black cross represents the median uncertainties. Both panels are on the same scale to make comparison easier.{\bf  Left (\ref{spt_teff}a):} CSGs from the SMC. {\bf Right (\ref{spt_teff}b):} CSGs from the LMC.}
\label{spt_teff}
\end{figure*}

The trend drawn by the LMC population alone appears to be nearly linear (see Fig.~\ref{spt_teff}b), because $r$ is higher than $r_{\rm s}$. It might be argued that the LMC correlation can arise from the presence of a few G~supergiants far from the core of the SpT distribution. The G~supergiants of the LMC do not seem to correspond to the same population as the K and M~supergiants, contrarily to the YSGs from the SMC \citepalias{dor16a}. Thus, there is a strong reason to study the correlation coefficients for the LMC without the G~supergiants. As there are only six G-stars in the LMC sample, the sample size is not significantly affected by removing them. We find that for the original sample, $r$ clearly decreases (see Table~\ref{corr_table}), although the correlation still remains highly significant. Interestingly, $r_{\rm s}$ is nearly unaltered. When Montecarlo calculated coefficients are considered, both r and rs decrease, although the difference between them becomes smaller. This behaviour is indicative of non-linearity ($r_{\rm s}\gtrsim r$), which in turn is not unexpected. The $T_{\rm eff}$ scale of \cite{lev2006} was indeed not linear, possessing different slopes for K and M~supergiants. The non-linearity seems less noticeable when G~supergiants are included, because they have a very high weight on $r$ due to their extreme positions in relation to the main cluster of points.

In the original SMC sample and in the combined MC sample, $r_{\rm s}$ is significantly higher than $r$. As we discussed earlier for the LMC, this indicates an underlying non-linearity in the $T_{\rm eff}$ scale. The Montecarlo-calculated coefficients, by contrast, are very similar. However, the fact that $r_{\rm s}$ and $r$ are equal is still indicative of non-linearity because $r_{\rm s}$ is less sensitive than $r$. The break in linearity for the LMC sample seems to be located around the change between K and M types \citep{lev2006}. For the SMC, instead, the trend between SpT and $T_{\rm eff}$ seems to change around K0 (see Fig.~\ref{spt_teff}a). Therefore, the slope of the $T_{\rm eff}$ scale appears to be different for G and K CSGs, as it seems to be different between K and M~supergiants. SMC CSGs present a significant dispersion in $T_{\rm eff}$ for types earlier than K1. However, all the points above the main trend (for $T_{\rm eff}>4\,500\:$K) correspond to low values of S/N (between 30 and 60). Thus, these points are not as trustworthy as those below them (see Fig.~\ref{errpar}).

In all the cases studied, the correlation coefficients derived show that the SpT sequence is strongly related to $T_{\rm eff}$. This result is in perfect agreement with the empirical results already presented in \citetalias{dor16a}, where we explored the relation between SpT and $T_{\rm eff}$ using EW(\ion{Ti}{i}) to connect them. In addition, we also found a significant but mild correlation between SpT and $M_{\rm bol}$ when mid- to high-luminosity CSGs (Ia and Iab) are considered \citepalias[see sects.~3.2 and~4.1.1 from][]{dor16a}. The correlation coefficients between SpT and $M_{\rm bol}$ for subsamples of stars brighter than a given value of $M_{\rm bol}$ lay between $-0.17$ and $-0.53$ \citepalias[see table~2 from][]{dor16a}.  The SpT\,--\,$T_{\rm eff}$ coefficients for the whole SMC and LMC samples (Table~\ref{corr_table}) are significantly higher than the SpT--$M_{\rm bol}$ coefficients found in \citetalias{dor16a}. If we apply the same cuts in luminosity used in \citetalias{dor16a} to explore the SpT--$M_{\rm bol}$ correlation, the correlation coefficients between $T_{\rm eff}$ and SpT do not vary significantly (Table~\ref{corr_table}). These results confirm the conclusions of \citetalias{dor16a}: SpT depends mainly on $T_{\rm eff}$, though it has a weaker, second-order relation to $M_{\rm bol}$.

\subsubsection{New temperature scales}
\label{temp_scale}

\begin{figure}
	\centering
	\includegraphics[trim=0.5cm 0.5cm 2cm 1.2cm,clip,width=8.5cm]{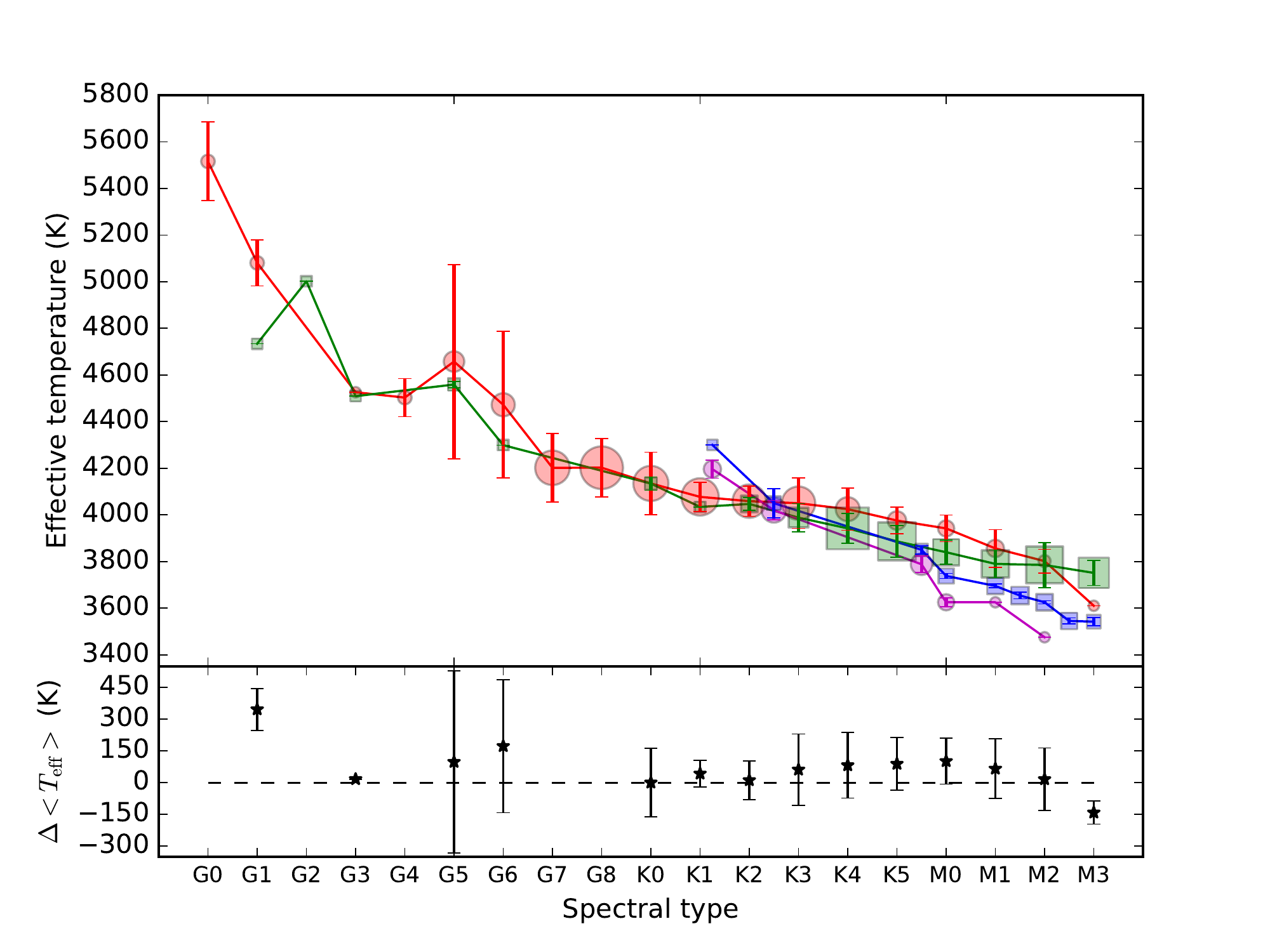} 
	\caption{{\bf Upper panel:} Effective temperature scale for our samples of CSGs from the SMC (red) and the LMC (green). We also overplotted the temperature scale derived by \citet{lev2006} for the SMC (purple) and the LMC (blue). The size of the symbols is proportional to the number of CSGs that contribute to each point. The absence of a symbol for a given SpT indicates that there are not CSGs with that SpT. The coloured error bars indicate the S/N-weighted standard deviation of the CSGs of the corresponding SpT. Note that for the SMC, the error bars lie inside the symbols in most cases. The exact values used for this figure are detailed in Table~\ref{tab_teff_scale}. {\bf Lower panel:} Differences (SMC$-$LMC) between the mean values of the effective temperature of each MC at each SpT. The differences have not been calculated for SpTs that are absent in one of the galaxies. The uncertainties are the sums of the corresponding standard deviations.}
\label{fig_teff_scale}
\end{figure}

The results found in Sect.~\ref{corr_spt_teff} imply a well-defined $T_{\rm eff}$ scale. Our next step is to describe this scale. \citet{lev2006} found that each galaxy seems to have its own temperature scale. In other words, RSGs from different galaxies having the same $T_{\rm eff}$ should present different SpTs. Therefore, each MC temperature scale must be considered separately. We derived an average $T_{\rm eff}$ and a standard deviation for each subtype, weighting each point by its corresponding S/N, within each galaxy. In this manner, we gave less weight to CSGs with larger uncertainties in $T_{\rm eff}$. We present all these values for each SpT in Table~\ref{tab_teff_scale} and in Fig.~\ref{fig_teff_scale}, to ease the comparison with fig.~9 of \cite{lev2006}.

\begin{table*}
	\caption{Effective temperature scale detailed for each subtype, for both MCs. For each SpT, the S/N weighted-mean of $T_{\rm eff}$ and its corresponding weighted standard deviation are shown (see Sect.~\ref{temp_scale} for further details) }
\centering
\begin{tabular}{c |  c c c |  c c c}
\hline\hline
\noalign{\smallskip}
&\multicolumn{3}{| c |}{SMC}&\multicolumn{3}{| c |}{LMC}\\
SpT&Number&<$T_{\rm eff}$> (K)&$\pm\sigma(T_{\rm eff})$ (K)&Number&<$T_{\rm eff}$> (K)&$\pm\sigma(T_{\rm eff})$ (K)\\
\hline
\noalign{\smallskip}
G0&3&5516&$\pm169$&--&--&--\\
G1&3&5081&$\pm99$&1&4734&--\\
G2&--&--&--&1&5002&--\\
G3&1&4526&--&1&4509&--\\
G4&3&4503&$\pm82$&--&--&--\\
G5&9&4657&$\pm417$&2&4559&$\pm14$\\
G6&12&4472&$\pm314$&1&4299&--\\
G7&29&4202&$\pm147$&--&--&--\\
G8&45&4202&$\pm125$&--&--&--\\
K0&30&4135&$\pm134$&2&4134&$\pm29$\\
K1&34&4077&$\pm62$&1&4034&--\\
K2&27&4059&$\pm65$&5&4047&$\pm27$\\
K3&27&4050&$\pm108$&8&3988&$\pm61$\\
K4&13&4024&$\pm91$&43&3942&$\pm64$\\
K5&7&3976&$\pm57$&35&3886&$\pm68$\\
M0&5&3942&$\pm57$&16&3840&$\pm52$\\
M1&6&3856&$\pm81$&17&3790&$\pm60$\\
M2&2&3802&$\pm51$&33&3785&$\pm97$\\
M3&1&3610&--&22&3751&$\pm54$\\
\hline
\end{tabular}
\label{tab_teff_scale}
\end{table*}

When we examine the typical $T_{\rm eff}$ for each SpT in each MC, we find that, in general terms, the values are compatible. For early G~stars, there are significant discrepancies between the typical $T_{\rm eff}$'s of each galaxy. However, the number of CSGs with these SpTs in both MCs is too low to consider this result as statistically meaningful. This applies as well to the subtype M3, where a significant difference is found, but  there is only one star of this type in the SMC, and its spectrum has a low S/N. We found small discrepancies in the typical $T_{\rm eff}$ of each galaxy for subtypes K5 and M0. These differences are significant at 1-$\sigma$, but not at the 2-$\sigma$ level. Moreover, they are derived from a low number of CSGs from the SMC. Since we find no significant differences between the typical $T_{\rm eff}$'s of adjacent SpTs (i.e.\ K3, K4, M1, or M2), we cannot interpret these small differences in K5 and M0 as an indication of two different SpT scales. Instead, our results are compatible with the existence of a unique $T_{\rm eff}$ scale followed by all CSGs, regardless of their metallicity. In this scenario, metallicity would only determine the typical SpTs, or in other words, which part of the $T_{\rm eff}$ scale is populated by the CSGs.

Using our data, the $T_{\rm eff}$ scales for the two MCs are not distinguishable (in statistical terms), which suggests a unique scale. However, this interpretation should be examined with care. The overlap between the two SpT distributions is very limited, and so any attempt to make a direct comparison between the LMC and SMC scales will be based on very few stars. We note that \cite{lev2006} studied the differences between CSGs from the Galaxy and the LMC, whose metallicities are different, without finding any significant differences between the corresponding $T_{\rm eff}$ scales, although the two galaxies have almost the same SpT coverage \citep[the difference is only half a subtype in the samples of][]{lev2006}. Moreover, we find a somewhat high dispersion of $T_{\rm eff}$ in each SpT. This is not unexpected. In addition to intrinsic uncertainties, there are many physical effects contributing to this dispersion: variations in luminosity by more than one order of magnitude within each subtype, implying large changes in the extension of the atmospheres; variations in mass-loss rate that affect the structure of the molecular envelope; variations in metallicity within each galaxy, and so on. The average value of the $T_{\rm eff}$ dispersions ($\sigma(T_{\rm eff})$), weighted by the number of stars in each SpT, is $100\:$K for the SMC and $48\:$K for the LMC. As the difference in $T_{\rm eff}$ between SpTs is only a few tens of Kelvin, for K and M types, it is likely that small samples would result in a non-meaningful scale due to the stochastic effects emerging from incomplete statistics.

Although we have observed almost all the stars used by \cite{lev2006}, we prefer not to compare their $T_{\rm eff}$'s individually because of the possibility of spectral variability, which might imply changes in $T_{\rm eff}$. Therefore, we simply compare our $T_{\rm eff}$ scale with those obtained by \cite{lev2006}. For late K to early M~CSGs, our scale covers a narrower range of $T_{\rm eff}$s when compared to \citet{lev2006}. Interestingly, while for early K types our scales seem fully compatible, below late~K we derived signficantly higher $T_{\rm eff}$'s. These results are only statistically significant for the LMC samples, as the number of M~RSGs from the SMC is almost negligible. This systematic difference is likely related to our different ways to calculate  $T_{\rm eff}$'s. While we obtained the $T_{\rm eff}$ from spectral synthesis of photospheric weak metallic lines, \cite{lev2006} employed the shape of their whole spectral range (from $4\,000$\AA{} to $9\,500$\AA{}), which is fully dominated by TiO bands for late K and M subtypes. As \cite{dav2013} demonstrated, using the TiO bands results in cooler temperatures than found when employing atomic lines. Finally, we point out that, although \citep{dav2013} did not find a $T_{\rm eff}$ scale, our temperatures for K and M CSGs fall within the $T_{\rm eff}$ range of $4\,100\pm150\:$K where they place all RSGs, except for the spectral types M2 and M3, which correspond to slightly lower temperatures. Our method uses atomic lines to calculate the $T_{\rm eff}$ and our sample has a high statistical weight. Therefore we can safely expect our $T_{\rm eff}$ scale to be a better representation of photospheric temperatures than those in previous works based on TiO bands.

\subsubsection{Spectral variability and effective temperature}
\label{spec_var}

\begin{figure*}
	\centering
        \includegraphics[trim=0.7cm 0.4cm 1.7cm 1.2cm,clip,width=8.5cm]{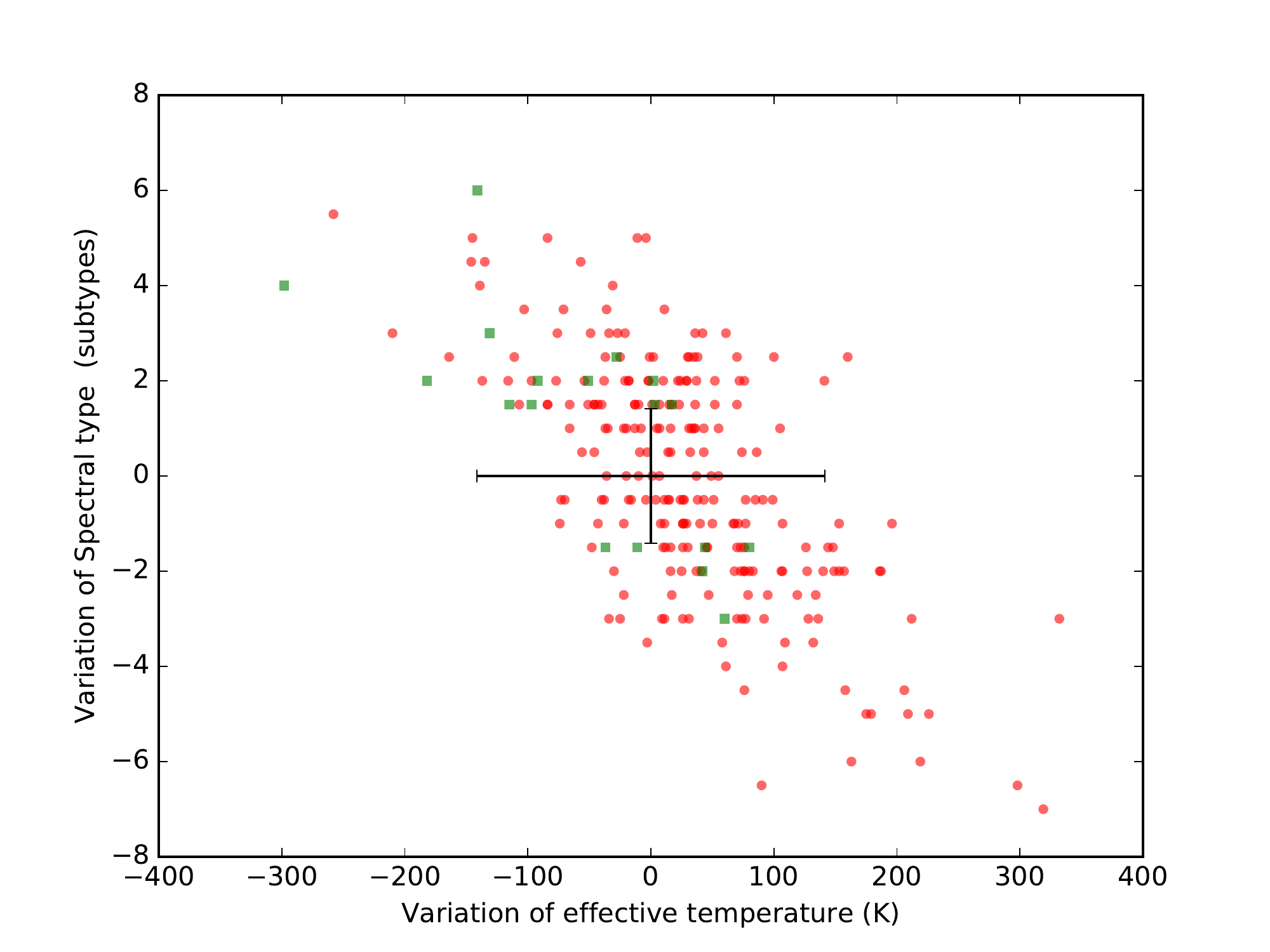}
	\includegraphics[trim=0.7cm 0.4cm 1.7cm 1.2cm,clip,width=8.5cm]{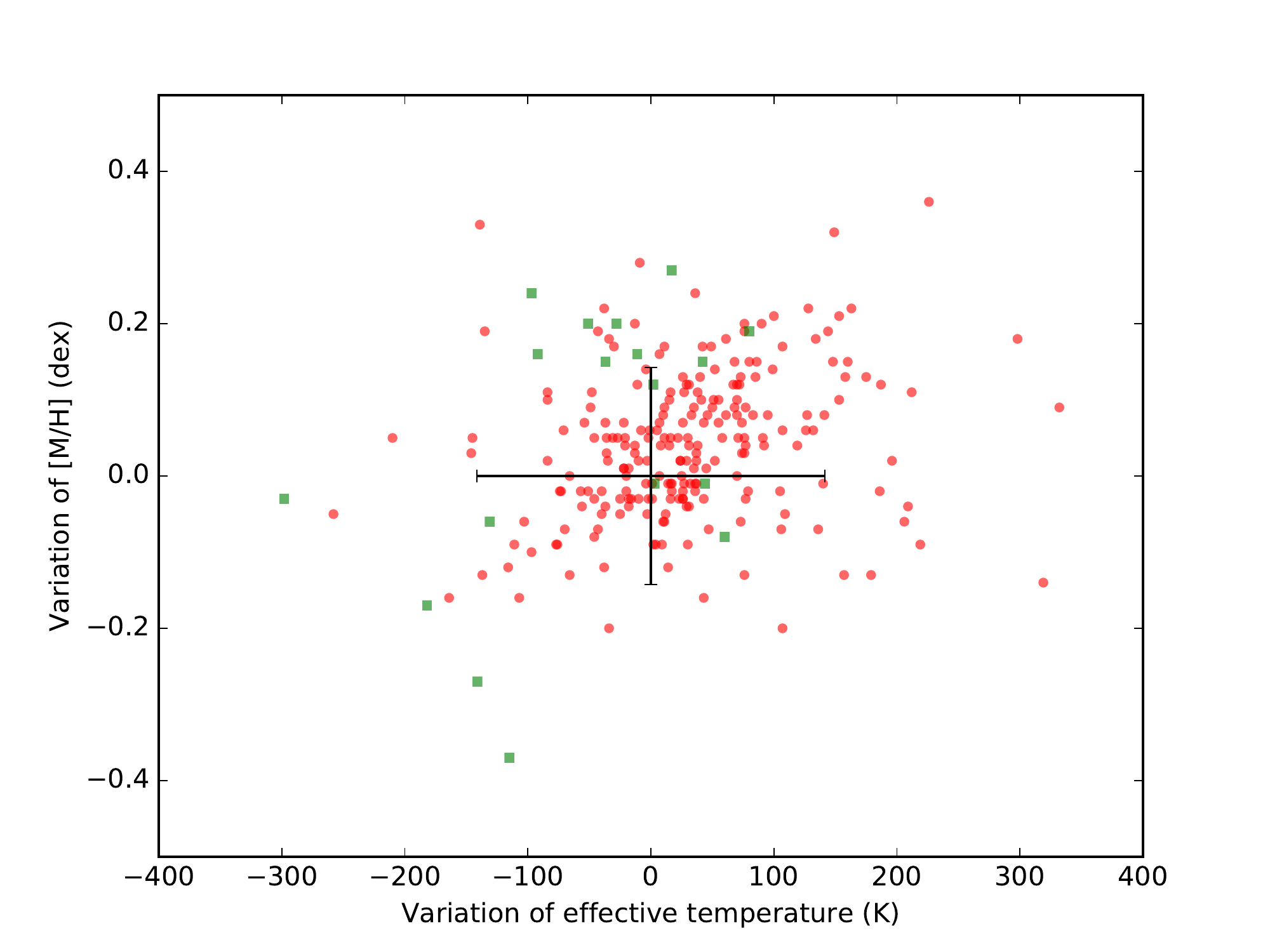} 
	\caption{Variations in the effective temperature against variations in {\bf Left (\ref{var_spt_teff}a):} SpT and {\bf Right (\ref{var_spt_teff}b)} [M/H]. Each point is the difference between two epochs for a given supergiant. The colour indicates the SpT that the CSG changed to. Green squares are LMC CSGs; red circles are SMC CSGs. The black cross at (0,0) shows the median error in each axis. Epochs when a star changed to SpTs later than M3 are not used.}
\label{var_spt_teff}
\end{figure*}

In \citetalias{dor16a} we analysed the spectral variability of our sample, using multi-epoch data for approximately a hundred stars from each MC that had showed significant changes in their spectral type ($\Delta_{\rm v}$SpT) between epochs. We found that spectral changes were accompanied by variations in spectral features generally assumed to respond to $T_{\mathrm{eff}}$, but not by variations in features that respond to luminosity. Now, we can test directly the relation between $\Delta_{\rm v}$SpT and $\Delta_{\rm v}T_{\rm eff}$. For this, we used the same procedure as in \citetalias{dor16a}: we took all the stars observed in more than one epoch and tagged as variable (i.e.,\ exhibiting a change larger than our typical error for SpT, one subtype, between any two given epochs) and computed differences in SpT and $T_{\rm eff}$ from one epoch to the other. When more than two epochs are available, we calculated the differences between all the possible epoch combinations (see Fig.~\ref{var_spt_teff}).

\begin{table}
\caption{Pearson ($r$) and Spearman ($r_{\rm s}$) coefficients obtained for the variables indicated. The coefficients marked as Montecarlo are the means and the standard deviations obtained from the $10\,000$ samples generated through a Montecarlo process (see text for details).}
\label{var_table}
\centering
\begin{tabular}{ c c | c c | c c }
\hline\hline
\noalign{\smallskip}
\multicolumn{2}{c|}{Variables}&\multicolumn{2}{c|}{From Montecarlo}&\multicolumn{2}{c}{Original sample}\\
X&Y&$r\pm\sigma_{\rm p}$&$r_{\rm s}\pm\sigma_{\rm s}$&$r$&$r_{\rm s}$\\
\noalign{\smallskip}
\hline
\noalign{\smallskip}
$\Delta_{\rm v}$SpT&$\Delta_{\rm v}T_{\rm eff}$&$-0.68\pm0.03$&$-0.65\pm0.04$&$-0.68$&$-0.62$\\
$\Delta_{\rm v}$SpT&$\Delta_{\rm v}[$M/H$]$ &$-0.04\pm0.06$&$-0.03\pm0.06$&$-0.04$&$-0.02$\\
$\Delta_{\rm v}$~$T_{\rm eff}$&$\Delta_{\rm v}[$M/H$]$ &$0.25\pm0.05$&$0.24\pm0.05$&$0.25$&$0.27$\\
\noalign{\smallskip}
\hline
\noalign{\smallskip}
\end{tabular}
\end{table}

As done for the correlation between SpT and $T_{\rm eff}$, we obtained the Pearson~($r$) and Spearman~($r_{\rm s}$) correlation coefficients for $\Delta_{\rm v}$SpT versus $\Delta_{\rm v}T_{\rm eff}$. Again, we did this for the original data and for $10\,000$ samples generated through a Montecarlo process, such as that explained in Sect.~\ref{corr_spt_teff}. The results of this process are shown in Table~\ref{var_table}. Both correlation coefficients, $r$ and $r_{\rm s}$, reveal the existence of a significant correlation. The significance is similar, though higher, than for the correlation obtained between $\Delta_{\rm v}$SpT and $\Delta_{\rm v}$EW(\ion{Ti}{i}) found in \citetalias{dor16a}, where the latter was used as proxy for $T_{\rm eff}$. As a sanity check, we also calculated the correlation coefficients of $\Delta_{\rm v}[$M/H$]$ with $\Delta_{v}$SpT and $\Delta_{\rm v}T_{\rm eff}$ (see Table~\ref{var_table}). In the case of $\Delta_{\rm v}[$M/H$]$/$\Delta_{v}$SpT, all correlation coefficients are compatible with zero, as expected (see Fig.~\ref{var_spt_teff}a). This not only confirms that our methodology does not introduce spurious correlations, but also provides an independent check that our calculations produce a non-skewed [M/H]. In the case of $\Delta_{\rm v}[$M/H$]$/$\Delta_{\rm v}T_{\rm eff}$ (see Fig.~\ref{var_spt_teff}b), we found a weak correlation that corresponds to the degeneracy between $T_{\rm eff}$ and [M/H]. Degeneracy between stellar parameters has been studied in the literature, as for example in \citet{sch14} and \citet{tin17}. In particular \citet{tin17} show very clearly the effects of parameter degeneracy in their figure 15. Thus, the degeneracy we found between $T_{\rm eff}$ and [M/H] is not unexpected at all. Indeed \citet{tin17} find that degeneracy between these two parameters in the same sense as found in this work is unavoidable and to a large degree independent of spectral resolution. Degeneration is an intrinsic issue to any calculation, especially for cool stars. It is posible to minimize its effects, as we have done in this work by a careful choise of our diagnostic features. We mixed low $\chi_{\rm l}$\footnote{Lower level excitation potential} lines, which will depend strongly on $T_{\rm eff}$, with some high $\chi_{l}$ lines that will strongly depend on [M/H].

 We thus find that not only there is a global correlation between SpT and $T_{\rm eff}$ for the sample as a whole, but also, when we consider the SpT changes of a given star, they correlate strongly to changes in $T_{\rm eff}$. Does this $\Delta_{\rm v}$SpT/$\Delta_{\rm v}T_{\rm eff}$ correlation imply that spectral variability is mainly driven by changes in T$_{\rm eff}$?. We know that other factors affect SpT, and \cite{dav2013} conclude that SpT in RSGs does not only depend on $T_{\rm eff}$. In view of the dispersion found in Fig.~\ref{spt_teff}, we cannot disagree with this conclusion. Indeed, \citetalias{dor16a} found a significant correlation between SpT and $M_{\rm bol}$ for the global sample, although this correlation is clearly weaker than that between SpT and $T_{\rm eff}$. However, when we consider the changes in SpT of a given star, it seems to imply only variations in $T_{\rm eff}$. The correlation of $\Delta_{\rm v}$SpT with the variations in lines sensitive to surface gravity (features classically used for luminosity classification) is nonexistant \citepalias{dor16a}. Moreover, despite the intrinsic degeneracy between $\Delta_{\rm v}$[M/H] with $\Delta_{\rm v}T_{\rm eff}$, the correlation between $\Delta_{\rm v}$SpT and $\Delta_{\rm v}$[M/H] is still compatible with 0, a result expectable if $\Delta_{\rm v}$SpT is not affected by spurious effects in our calculations. Altough there could be other physical properties, beyond surface gravity, that may have a significant impact on $\Delta_{\rm v}$SpT, but they are unlikely to be the main driver of changes in SpT. In view of all this, even though the SpT of an RSG does not only depend on $T_{\rm eff}$, we have to conclude that spectral variations are, to a large extent, driven by changes in $T_{\rm eff}$, regardless of other second order effects. This conclusion reinforces the idea advanced in Sect.~\ref{corr_spt_teff}: although SpT is sensitive to luminosity, it is a second order effect when compared to the dependence on $T_{\rm eff}$. The fact that this dependence on luminosity is seen when we consider the whole sample, but does not appear when we study the variations of a given star supports the idea that it is related to evolutionary process, as suggested by \citet{bea2016}.

\subsubsection{The typical temperatures of cool supergiants}
\label{temp_scale}

In previous sections we have shown that a strong correlation exists between SpT and $T_{\rm eff}$. In addition, we have failed to find a statistically significant difference between the $T_{\rm eff}$ scales of the LMC and the SMC. On the other hand, the two MCs have different typical metallicities and different SpT distributions, as we already showed in \citetalias{dor16a}. Here we want to test if there is indeed a significant difference between the $T_{\rm eff}$ distributions from different environments. 

\begin{table}
\caption{Fraction of rejections of the null hypothesis of the KST (at $\alpha=0.02$) done over pairs of $T_{\rm eff}$ distributions from each MC, for different-size subsamples. The rejection of the null hypothesis means that the samples of each MC are considered as significantly different. The fractions were calculated over groups of $10\,000$ subsamples randomly taken. The size refers to the nomber of targets in the subsample of each MC. We provide as uncertainties of our fractions the 2-$\sigma$ confidence intervals, which are equal to $1/\sqrt[]{n}$, where $n=10\,000$. The limited tag refers to our sample restricted to RSGs (K and M SpTs) of mid to high luminosity (Ia to Iab). See text for details.}
\label{KST_table}
\centering
\begin{tabular}{ c c c }
\hline\hline
\noalign{\smallskip}
Size of&\multicolumn{2}{c}{Fraction of rejections of the null hypothesis}\\
the sample&Not limited&Limited\\
\noalign{\smallskip}
\hline
\noalign{\smallskip}
5&$0.00\pm0.01$&$0.00\pm0.01$\\
7&$0.19\pm0.01$&$0.12\pm0.01$\\
10&$0.30\pm0.01$&$0.19\pm0.01$\\
15&$0.71\pm0.01$&$0.54\pm0.01$\\
20&$0.96\pm0.01$&$0.89\pm0.01$\\
25&$0.99\pm0.01$&$0.95\pm0.01$\\
50&$1.00\pm0.01$&$1.00\pm0.01$\\
\noalign{\smallskip}
\hline
\noalign{\smallskip}
\end{tabular}
\end{table}

We compared our samples from the LMC and the SMC using a Kolmogorov-Smirnov Test (KST). We required a level of significance $\alpha=0.02$ to reject the null hypothesis (both samples are indistinguishable), finding that the KST rejected the null hypothesis for our samples by a wide margin. For stronger confirmation, we calculated through a Montecarlo process $10\,000$ random samples of the size of our original sample for each galaxy, by allowing each star to take random values of its effective temperature according to a gaussian distribution. After this, we passed pairs of these samples (one from each galaxy) through the KST. The null hypothesis was rejected for all cases. Under the light of this result, we can conclude that the $T_{\rm eff}$ distribution of CSGs is significantly different between the two galaxies.

Our results appear to be contradictory with conclusions of \cite{dav2013}, who argued that all CSGs have roughly the same $T_{\rm eff}$. They based this conclusion on the fact that they could not find significant differences between the typical $T_{\rm eff}$ of their RSG samples from the two MCs. Successive works by the same group have supported this conclusion for other CSG populations. At this point, we have to remark that our results do not imply in any manner that the values of $T_{\rm eff}$ calculated through their methods are incorrect. In fact, we have showed in Sect.~\ref{lte} that the values obtained through our methods are statistically indistinguishable from theirs, at least for the small sample that we could use for the comparison. Under the light of our results, we can speculate that their conclusion, as well as succesive confirmations given in later works, is a consequence of two effects combined: the limited size of their samples and the dispersion in $T_{\rm eff}$ that we find within each subtype (see Sect.~\ref{temp_scale} and Fig.~\ref{tab_teff_scale}). In order to confirm this assertion, we devised the following test: we randomly took subsamples of different sizes. For each different subsample size, we took $10\,000$ random subsamples from each MC. Then, we applied the KST (at $\alpha=0.02$) to each of the $10\,000$ pairs of subsamples (one from each galaxy). Finally, we calculated the fraction of the $10\,000$ pairs of subsamples of a given size for which the null hypothesis (both subsamples are statistically indistinguishable) was rejected. Results are provided in Table~\ref{KST_table}; they clearly indicate that, whenever small subsamples are taken, the null hypothesis cannot be rejected with confidence. As the sample size increases, the fraction of rejection increases. By the time a sample size reaches 50~CSGs, from each galaxy, the null hypothesis is rejected in all experiments. Contrariwise, with the sample sizes of \cite{dav2015}, the null hypothesis cannot be rejected in most experiments.

It could be argued that our samples are not fully equivalent to those of \cite{dav2015}, because we include a large number of G~stars and also low luminosity CSGs (Ib and Ib\,--\,II). Therefore, we repeated our calculations using only RSGs (K and M types) of mid to high luminosity (Ia and Iab luminosity classes). This restriction has the added advantage that we also avoid the systematic difference between the LMC and SMC samples at low luminosities, caused by the difference in distance modulus ($0.5\:$mag) between the two galaxies. As can be seen in Table~\ref{KST_table}, the rejections become even lower for small samples. However, as in the non-limited subsamples case, for sizes of 50~CSGs the rejections reach 100\%.

\subsection{Metallicity effects}
\label{metal_eff}

In the previous section we concluded that CSGs in the SMC present a significantly different $T_{\rm eff}$ distribution from those in the LMC. This difference, also reflected in their SpT distribution, is believed to be a consequence of the difference in typical metallicity \citep{eli1985}. In fact, evolutionary models \citep{eks2012,geo2013} predict a shift in the lowest $T_{\rm eff}$ that CSGs can reach during their evolution due to metallicity. Previous works obtained contradictory results in this matter. On one hand, \cite{dro2012} analysed 189~RSGs from the galaxy M33. They found that the $T_{\rm eff}$  distribution of RSGs seems to change for different galactocentric radius \citep[see fig.~22 of][]{dro2012}, which implies different typical metallicities. On the other hand, \cite{gaz2015} studied a sample of 27~RSGs from the galaxy NGC~300. Although they obtained a clear trend between the galactocentric distance of the RSGs and their metallicity (see their fig.~6), in good agreement with previous works, they did not find a clear correlation between metallicity and $T_{\rm eff}$ (see their fig.~7).

\begin{figure*}
	\centering
	\includegraphics[trim=0.5cm 0.4cm 1.8cm 1.2cm,clip,width=8.8cm]{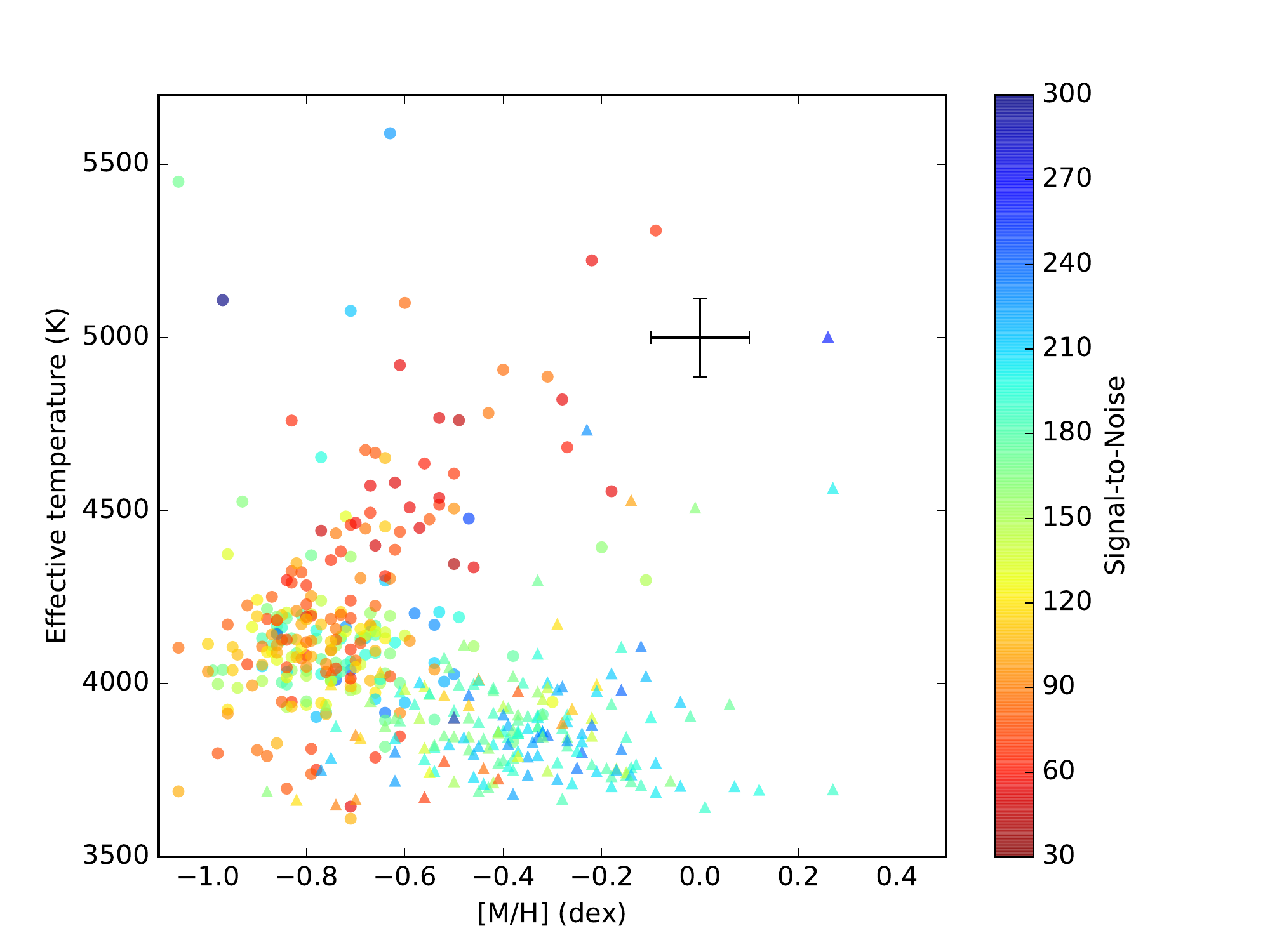}
	\includegraphics[trim=0.5cm 0.4cm 1.8cm 1.2cm,clip,width=8.8cm]{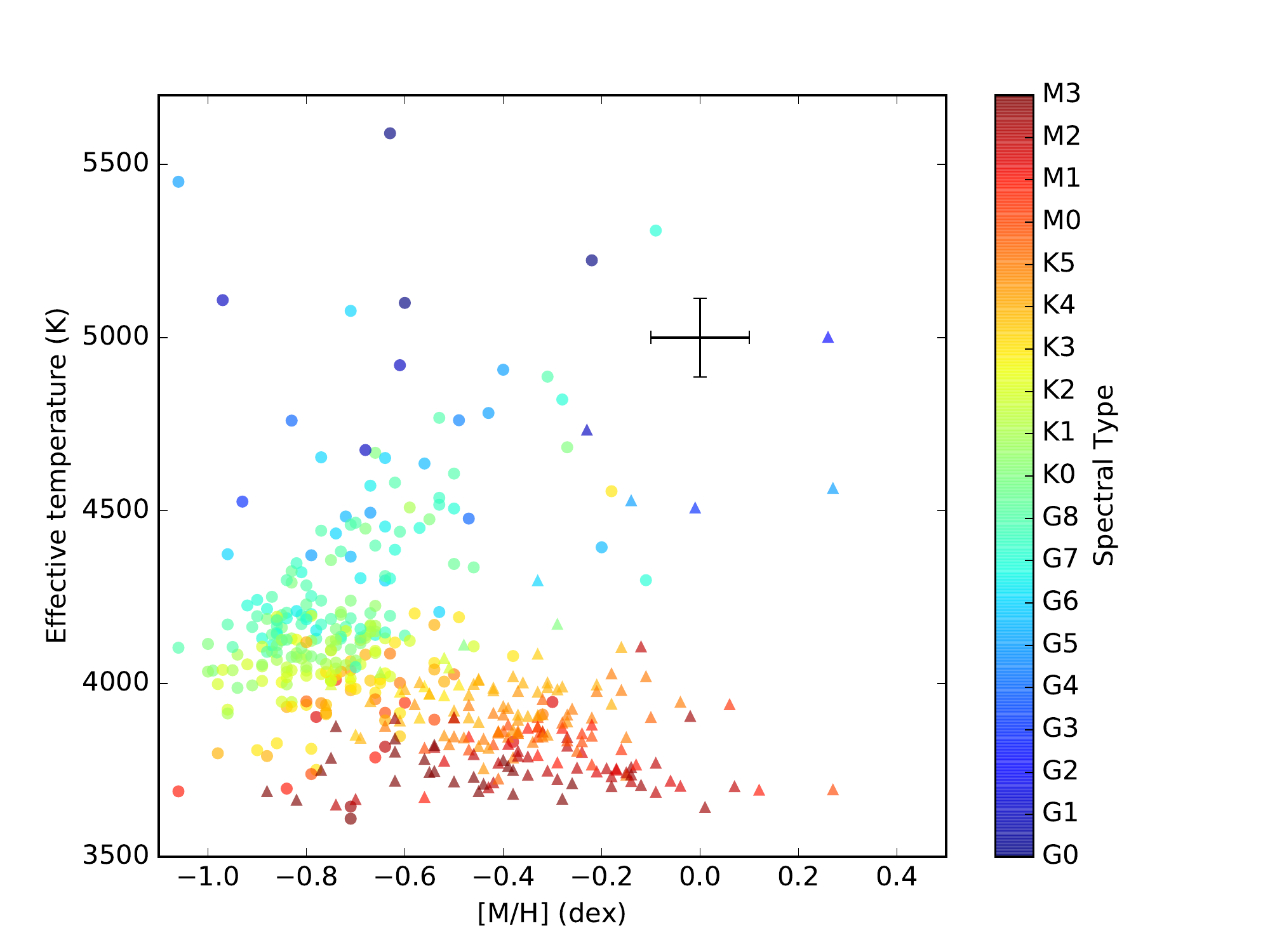} 
        \caption{Effective temperature against Metallicity for CSGs from both, the SMC (circles) and the LMC (triangles). The black cross represents the median uncertainties. {\bf Left (\ref{teff_z}a):} The color indicates the S/N. {\bf Right (\ref{teff_z}b):} The color indicates the SpT.
        }
\label{teff_z}
\end{figure*}

\begin{table*}
	\caption{Pearson ($r$) and Spearman ($r_{\rm s}$) coefficients obtained for the correlations between $T_{\rm eff}$ and [M/H] in different subsamples from both MCs. Some subsamples were limited to S/N >100 or to include only luminous RSGs (those having K or M types and Ia or Iab luminosity classes). The coefficients tagged as Montecarlo are the means and the standard deviations obtained from the $10\,000$ samples generated through Montecarlo (see text for details). }
\label{Z_Teff_table}
\centering
\begin{tabular}{c | c c | c c | c}
\hline\hline
\noalign{\smallskip}
&\multicolumn{2}{c|}{From Montecarlo}&\multicolumn{2}{c}{Original sample}&Size of\\
Sample&$r\pm\sigma_{\rm p}$&$r_{\rm s}\pm\sigma_{\rm s}$&$r$&$r_{\rm s}$&the sample\\
\noalign{\smallskip}
\hline
\noalign{\smallskip}
SMC&$0.35\pm0.05$&$0.33\pm0.06$&$0.35$&$0.25$&257\\
LMC&$0.18\pm0.06$&$0.11\pm0.06$&$0.18$&$-0.06$&188\\
SMC (S/N>100)&$-0.04\pm0.07$&$-0.03\pm0.07$&$-0.04$&$-0.02$&155\\
LMC (S/N>100)&$0.15\pm0.06$&$0.08\pm0.07$&$0.15$&$-0.11$&175\\
SMC (RSGs)&$0.07\pm0.07$&$0.07\pm0.08$&$0.07$&$0.08$&137\\
LMC (RSGs)&$-0.19\pm0.08$&$-0.18\pm0.07$&$-0.19$&$-0.17$&152\\
\noalign{\smallskip}
\hline
\noalign{\smallskip}
\end{tabular}
\end{table*}

In Fig.~\ref{teff_z}, we represent our values of $T_{\rm eff}$ and [M/H] for all the stars in our sample. As can be seen, the distribution of CSGs in each MC occupies different positions in the diagram. However, there is no clear correlation within the sample of each galaxy. To test statistically if there is a noticeable trend in the [M/H] span within each galaxy, we calculated the correlation coefficients ($r$ and $r_{\rm s}$). We used the same methodology as in Sect.~\ref{corr_spt_teff}. The results are shown in Table~\ref{Z_Teff_table}.

The coefficients for the LMC are very low, close to non-significant. Instead, for the SMC there is a weak correlation. However, in both cases the sign is the opposite to that expected, in the sense that we find higher $T_{\rm eff}$'s at higher metallicities. This result is a direct consequence of the presence of G~CSGs. In the case of the LMC, as explained above, these stars are not part of the same population as RSGs. For the SMC, the problem is the presence of many G~supergiants with low S/N, as can be seen in Fig.~\ref{teff_z}a. To avoid this last effect, we repeated all our calculations imposing a lower limit on S/N$\geq100$. The results are also shown in Table~\ref{Z_Teff_table}. A weak correlation persists for the LMC objects, because its G~CSGs present mid to high S/N, but in the case of the SMC the correlation simply disappears. Finally, we wanted to calculate again the correlation for the LMC but only including K and M~RSGs. Moreover, to allow a more direct comparison with the works of \cite{dro2012} and \cite{gaz2015}, we calculated the correlation only for the subsample of luminous RSGs (i.e.\ K and M types only and luminosity classes Ia and Iab only). With these limits, the LMC presents a weak correlation with the expected sign. Instead, for the SMC there is not a significant correlation, as in the case of the cut in S/N because the G~CSGs were most of the stars with low S/N.

From all these results, we conclude that there is not a noticeable correlation between [M/H] and $T_{\rm eff}$ among the CSGs of each galaxy. This result has an interesting contrast with the fact that CSGs from the SMC are less metallic than those from the LMC, but also significantly cooler. We can advance two possible reasons why we do not find any significant correlation within each galaxy. Firstly, we found average [M/H]$_{\rm LMC}=-0.33\pm0.20\:$dex and [M/H]$_{\rm SMC}=-0.73\pm0.17\:$dex. The standard deviation within each population is not very high ($\approx0.2\:$dex) when compared to the difference [M/H]$_{\rm LMC}-$\,[M/H]$_{\rm SMC}$ ($0.40$~dex). Secondly, the existence of significant spectral variability (especially among the SMC sample) can blur away any correlation, because a star of a given [M/H] may present very different $T_{\rm eff}$'s depending on the observation moment. As variability is more frequent in the SMC sample, this effect may explain why we find a weak correlation for the LMC sample, but not for that from the SMC. Under these assumptions the results of \cite{gaz2015} are not entirely unexpected. Although their sample spans a range of $0.6\:$dex, its standard deviation is only $0.14\:$dex, i.e.\ even smaller than the dispersions of our MC samples. Unfortunately, \cite{dro2012} do not provide values of metallicity for their sample. Thus, a similar analysis is not possible for their work. 

The idea that the typical SpT of a population is related to its metallicity stems from the comparative study of a few populations \citep{lev2012}. Our results seem to confirm warmer temperatures at lower [M/H], as the population in the SMC is clearly warmer than that in the LMC. However, the idea that there is a smooth functional relation between typical $T_{\rm{eff}}$ and [M/H] is an unproven extrapolation, perhaps suggested by the behaviour of theoretical tracks. Our results do not provide evidence in this sense.  In fact, despite the significant difference in [M/H] between the RSGs from the LMC and those from our Galaxy, the average SpT difference appears to be at most one subtype \citep{eli1985,lev2013}, which is almost unnoticeable. Therefore, it cannot be translated into a very large difference in $T_{\rm eff}$. For example, in \cite{lev2006} the difference between the LMC and the Milky Way is not significant at all (only $15\:$K). In view of this, the existence of a smooth correlation between the typical temperature of CSGs and metallicity must be kept on hold until further tests can be performed.

\subsection{Comparison with evolutionary tracks}

\begin{figure*}
	\centering
	\includegraphics[trim=1cm 0.4cm 1.8cm 1.2cm,clip,width=8.8cm]{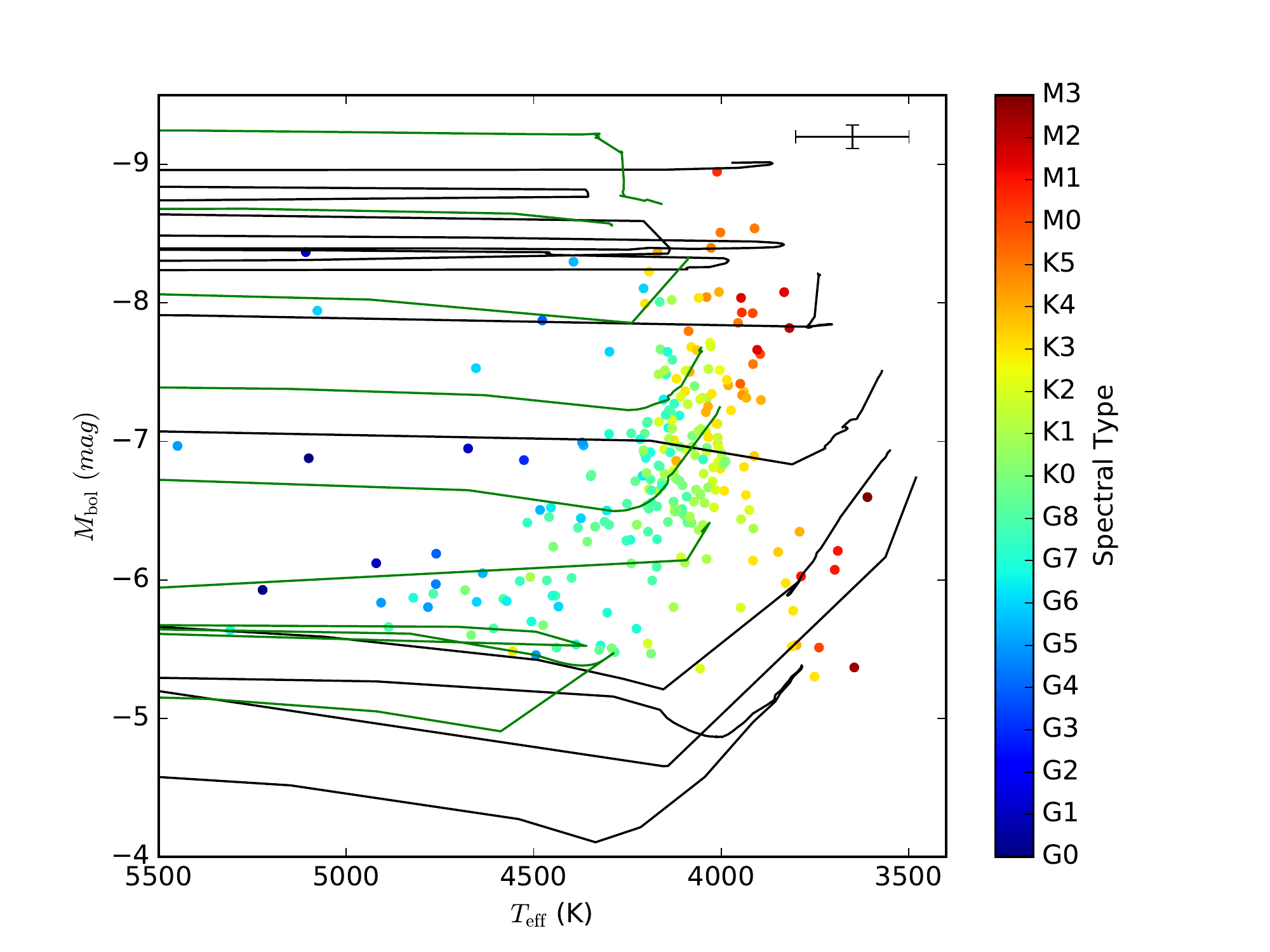}
	\includegraphics[trim=1cm 0.4cm 1.8cm 1.2cm,clip,width=8.8cm]{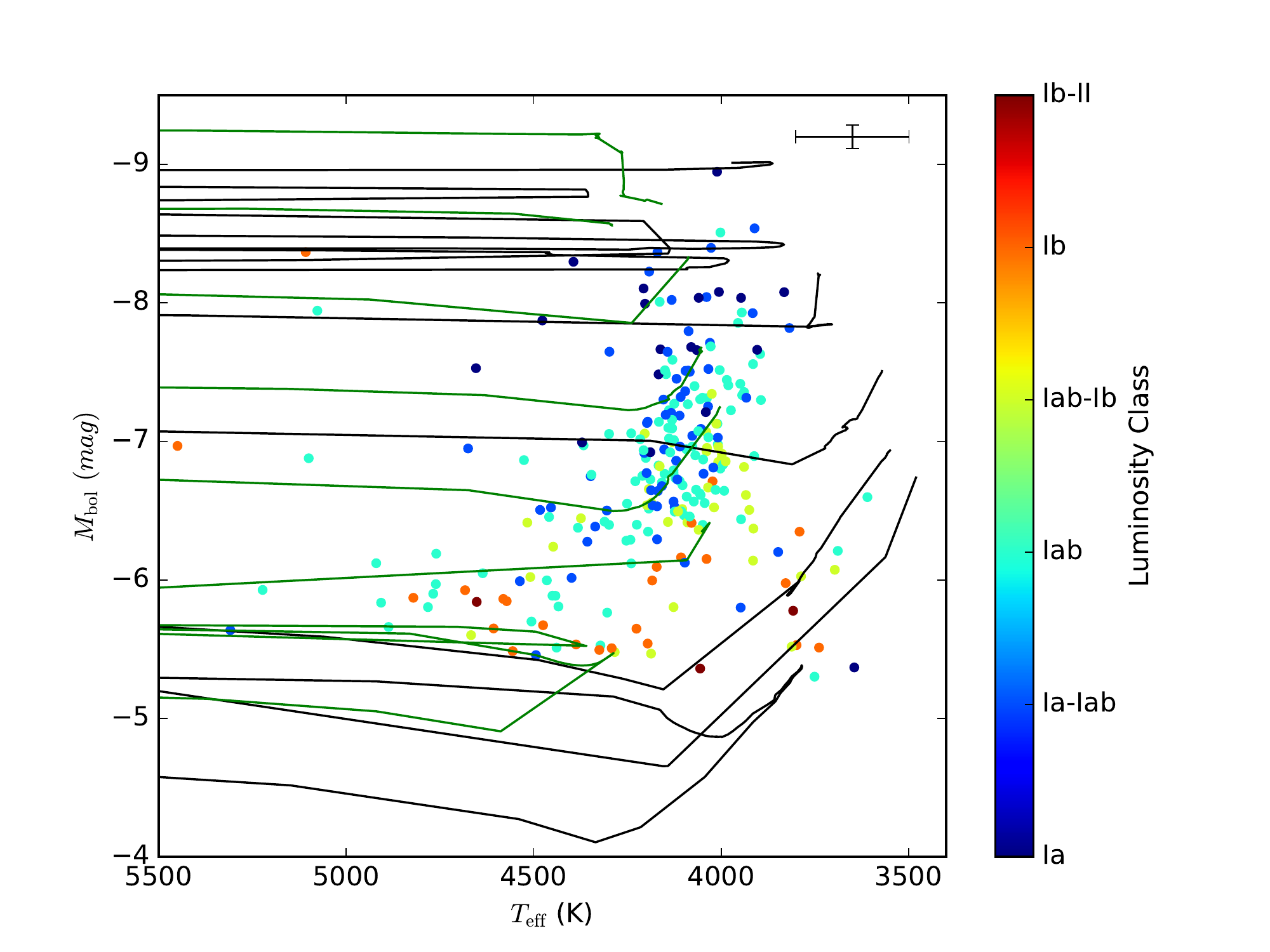}
        \caption{Our sample of CSGs from the SMC, plotted together with Geneva evolutionary tracks. Green tracks are for the typical metallicity of the SMC \citep{geo2013}. Black tracks are for Solar metallicity \citep{eks2012}. The tracks shown correspond (for both metallicities), from bottom to top, to stars of 9, 12, 15, 20, 25, and $32\:M_{\sun}$. The black cross represents the median uncertainties.
        {\bf Left (\ref{HR_SMC}a):} The color indicates the spectral types of the sample.
        {\bf Right (\ref{HR_SMC}b):} The color indicates the Luminosity classes of the sample.
        }
\label{HR_SMC}
\end{figure*}

\begin{figure*}
	\centering
	\includegraphics[trim=1cm 0.4cm 1.8cm 1.2cm,clip,width=8.8cm]{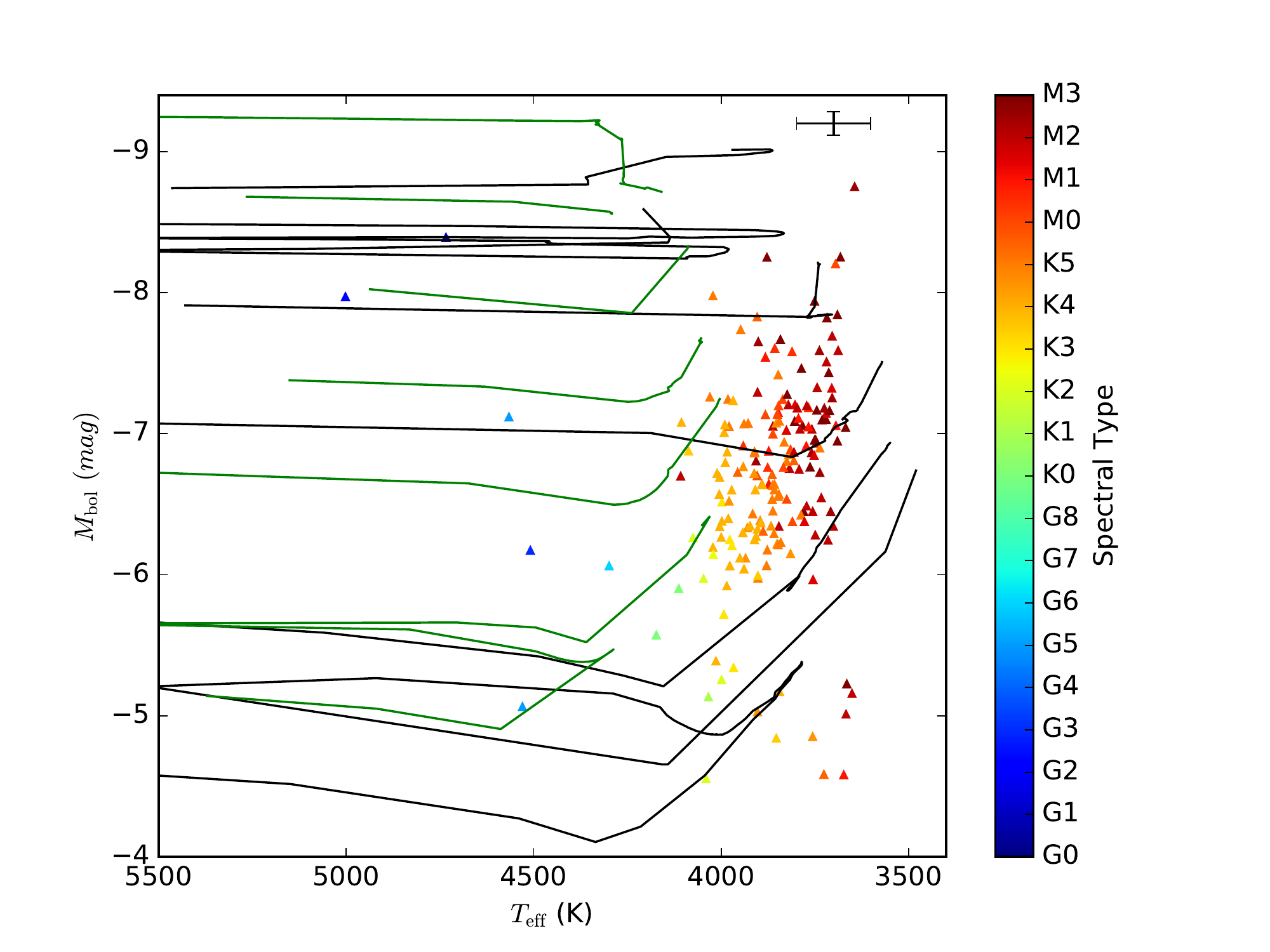} 
	\includegraphics[trim=1cm 0.4cm 1.8cm 1.2cm,clip,width=8.8cm]{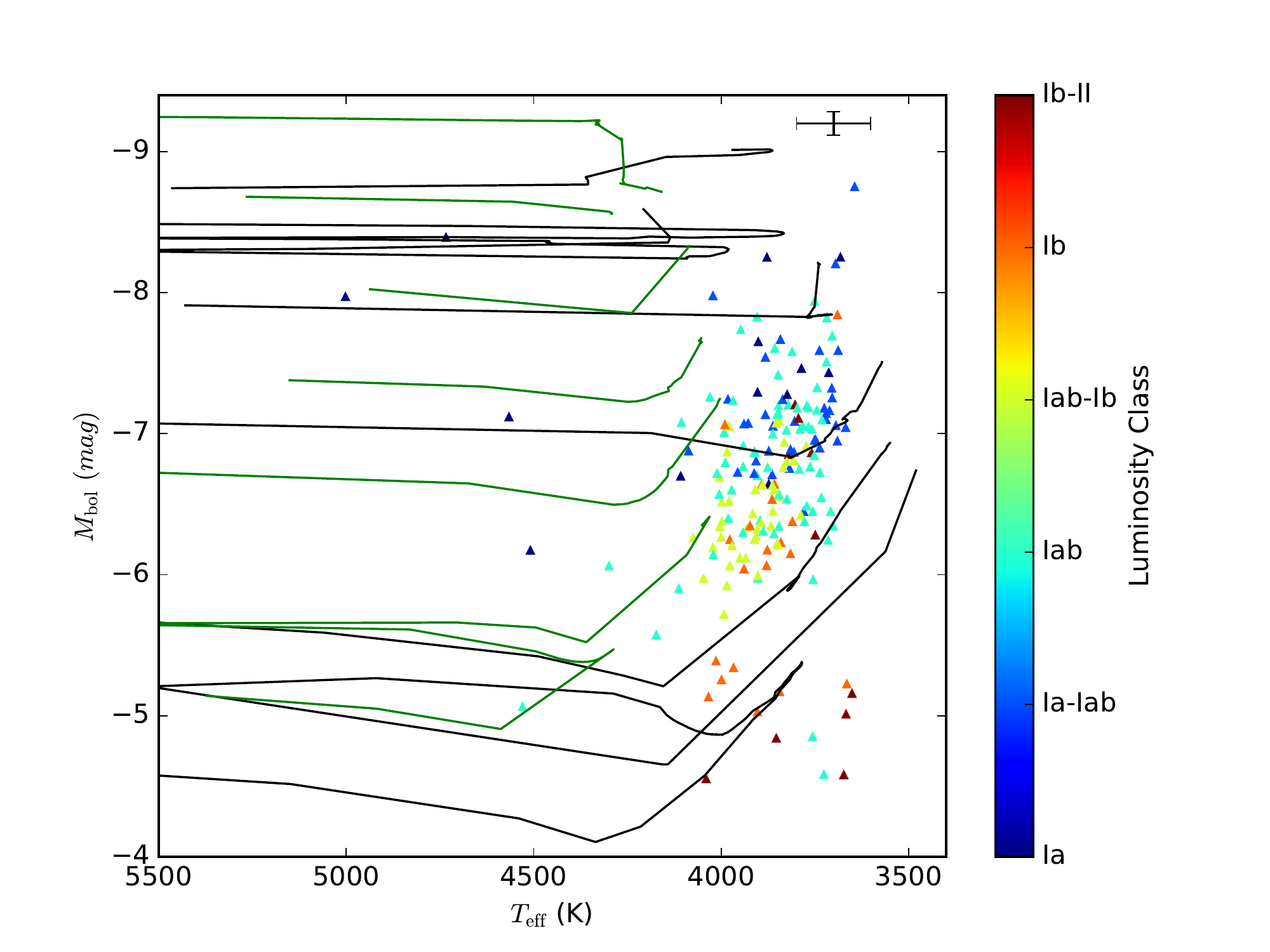} 
        \caption{Our sample of CSGs from the LMC (excluding objects with spectral type M3 and later), plotted together with Geneva evolutionary tracks. Symbols are as in Fig.~\ref{HR_SMC}. No tracks for LMC metallicty are available. Note that the sharp cut at low temperatures is almost certainly an artefact of our sample selection and procedure.
        {\bf Left (\ref{HR_LMC}a):} The color indicates the spectral types of the sample.
        {\bf Right (\ref{HR_LMC}b):} The color indicates the Luminosity classes of the sample.
        }
\label{HR_LMC}
\end{figure*}

We calculated absolute bolometric magnitudes for our CSGs in \citetalias{dor16a}. Using the $T_{\rm eff}$'s calculated in this work, we can plot our CSGs in the Hertzsprung-Russel diagram. In Figs.~\ref{HR_SMC} and~\ref{HR_LMC} we compare our data to Geneva evolutionary tracks. We used them to illustrate the state-of-the-art theoretical evolutionarty tracks because they are widely employed within the literature. Although there are other tracks available \citep[e.g.][]{bro2011}, they are qualitatively similar to those of Geneva and the range of $T_{\rm eff}$'s covered is also very similar \citepalias[see][]{dor16a}. 

Unfortunately, Geneva tracks for the LMC typical metallicity are not available. Instead, we provide the evolutionary tracks for solar metallicity ($Z=0.014$) to help in the interpretation of the figures. The evolutionary tracks for the LMC typical metallicity should fall between those for SMC and Solar metallicities. 

Firstly, we must remark that the samples plotted are biased towards hot temperatures, because all stars with types later than M3 have been discarded from our analysis. Given the SpT--$T_{\rm eff}$ correlation we found, we should expect them to have lower temperatures. In addition, several CSGs from our sample were discarded because they reach the edges of our synthetic spectra grids (see Sect.~\ref{fin_err_bud}). Therefore, there should be many more CSGs with lower temperatures than those shown, specially in the case of the LMC. Any interpretation of the figures must take this bias into consideration. Secondly, for clarity we do not represent tracks with rotation. The main effect of rotation is an increase in the predicted luminosity for each track. For most, it does not affect substantially the lowest temperature reachable. The change is only noticeable for the tracks with the highest masses, where we do not have almost any CSG. Therefore, the effect of rotation does not impact our analysis. 

In the case of the SMC (Fig.~\ref{HR_SMC}), the overall agreement with our results and the evolutionary tracks is good. CSGs with luminosities between $M_{\rm bol}\sim-6.5\:$mag and $M_{\rm bol}\sim-7.5\:$mag present temperatures compatible with the evolutionary tracks. This group represents most of the SMC sample. However, there are two groups of CSGs with $T_{\rm eff}<4000\:$K that are clearly colder than any $T_{\rm eff}$ shown in the evolutionary tracks at any given luminosity for the SMC. One group is formed by CSGs less luminous than $M_{\rm bol}\sim-6.5\:$mag. Some of these CSGs present the lowest temperatures in the SMC sample. In fact, these low luminosity CSGs reach temperatures as low as those predicted for Solar metallicity. The other group are stars brighter than $M_{\rm bol}\sim-7.5\:$mag with moderately low $T_{\rm eff}$. The presence of these highly luminous CSGs, with temperatures lower than those predicted, was noticed in previous works \citep[e.g.][]{lev2007,dav2013}. As the lowest $T_{\rm eff}$ reachable by a CSG is theoretically connected to its metallicity, we checked the metallicities of these stars. The [M/H] of all but three is inside the typical values of the SMC sample. Thus, it seems that metallicity cannot explain their lower temperatures. Furthermore, \citet{dav2013} discussed the assumptions made for treatment of convection used to generate the evolutionary tracks. Interestingly, they mention that the locations of these cool CSGs could be moved to cooler or warmer temperatures if the assumptions regarding convection are modified.

For the LMC (Fig.~\ref{HR_LMC}), the comparison is more problematic. Geneva evolutionary tracks of high-mass stars are not available for the typical [M/H] of the LMC. Consequently, we can only employ the tracks for SMC and solar metallicities to evaluate the behaviour of stars. Most LMC CSGs have $T_{\rm eff}$ values within the range displayed by the evolutionary tracks. However, many stars, especially among those more luminous than $M_{\rm bol}\lesssim-7\:$mag present $T_{\rm eff}$ values expected for solar metallicity, or even slightly lower. Again, those stars are not much more metallic than the typical values of our LMC sample. Moreover, as we explained above, we must take into account that there must be many RSGs with even lower $T_{\rm eff}$'s that are not included in our plot.

From our analysis of both MCs, we found that the agreement between evolutionary tracks and our sample is, in general terms, good. However, there are CSGs with lower temperatures than those present in the evolutionary tracks. Some of them are low luminosity CSGs and seem to correspond to the evolutionary track of $9\:M_{\sun}$. Therefore, although their spectral morphologies correspond to supergiants, they may be in fact highly luminous red giants. During the AGB phase, hot bottom burning may increase the luminosity of massive giants even above the classical limit of $M_{\rm{bol}}=-7.1$, thus mimicking the behaviour of more massive stars \citep[e.g.][]{ventura00}  However, we can say very little about this group, given that our sample is highly incomplete at such low luminosities.

Most of the CSGs with $T_{\rm eff }$ lower than those of evolutionary tracks in both MCs are concentrated at high luminosities. These CSGs have especially late spectral types and, as can be seen in fig.~19 of \citetalias{dor16a}, they present the highest mass-loss rates in both galaxies. The luminosity of these CSGs is compatible with evolutionary tracks for $\geq20\:M_{\sun}$. It is tempting to think that these high-mass tracks have a problem with their minimum $T_{\rm eff}$. Moreover, the evolutionary tracks for these high masses predict that CSGs do not stay at the coolest edge of the track for long, but instead return quickly back to warmer $T_{\rm eff}$. However, there is another possibility, which we already advanced in \citetalias{dor16a}. Mid-luminosity CSGs might perhaps evolve towards higher luminosities and lower $T_{\rm eff}$'s than predicted by $\sim12$\,--$15\:$M$_{\sun}$ evolutionary tracks. If so, they would form a group of CSGs with luminosities typical of higher initial masses ($M_{\rm bol}\lesssim-7\:$mag). Recent results by \citet{bea2016} support this scenario. They found that, while most RSGs in the LMC cluster NGC~2100 present luminosities between $\log(L/L_{\sun})\sim4.5$ ($M_{\rm bol}\sim-6.5\:$mag) and $\log(L/L_{\sun})\sim4.8$ ($M_{\rm bol}\sim-7.3\:$mag), there are two RSGs with significantly higher luminosities: $\log(L/L_{\sun})>4.95$ (M$_{\rm bol}\lesssim-7.6\:$mag). Unfortunately, as our sample does not represent a coeval population we cannot verify this scenario. Even then, our data show that there is a substantial number of CSGs with stellar atmospheric parameters inconsistent with current evolutionary tracks. Therefore, it is necessary to find an explanation for such high-luminosity CSGs inside the evolutionary theory of high-mass stars.

\section{Conclusions}
\label{conclus}

In this work, we have obtained stellar atmospheric parameters ($T_{\rm eff}$ and [M/H]) for the largest homogeneous sample of CSGs belonging to the MCs. For this, we used spectral features found in the CaT spectral range observed at mid-resolution. The results of the present manuscript are summarized below:

\begin{enumerate}

\item We compare our results to NLTE calculations using the sample observed by \citet{dav2015}, and find a small systematic difference. This difference is not significant given our internal uncertainties, but this result should be contrasted with a larger sample.

\item We find a significant correlation between $T_{\rm eff}$ and SpT. This correlation is much stronger than that between SpT and $M_{\rm bol}$. We also find a significant correlation when we analyse the variations of SpT and $T_{\rm eff}$. Therefore, we conclude that SpT is mainly driven by $T_{\rm eff}$, although luminosity has some influence on it.

\item From the correlations described above, we conclude that there is a meaningful temperature scale for CSGs. This scale seems to be the same for stars in the two MCs. At least, given our level of precision, we cannot tell any difference between the scales found for the two MCs. Although more studies are needed, our results seem to indicate that all CSGs belonging to the MCs are on a similar $T_{\rm eff}$ scale.

\item Although the samples of both MCs seem to follow the same $T_{\rm eff}$ scale, we report a significantly  different range of $T_{\rm eff}$ for each MC. This together with conclusion (ii), implies that the shift in the typical SpT of each galaxy is caused by the difference in $T_{\rm eff}$.

\item When we study the relation between [M/H] and $T_{\rm eff}$, we find that for the sample from the LMC, there is a weak correlation, and for that from the SMC there is no correlation at all. From these and other statistical analyses, we conclude that the relation between $T_{\rm eff}$ and [M/H] is subtle, and unlikely linear. In consequence, the effect of [M/H] on the $T_{\rm eff}$ distribution is only noticeable for samples with very different typical metallicities, as when CSG samples from the SMC and the LMC are compared to each other.

\item We compare our samples with Geneva evolutionary tracks. We find that the physical properties of mid-luminosity CSGs from both MCs, which represent the majority of our samples, match well the evolutionary tracks. However, most of the high-luminosity CSGs present temperatures significantly lower than expected from the evolutionary tracks. This result indicates that evolutionary tracks are not predicting correctly the nature of high-luminosity RSGs. Therefore, these stars require further research, both theoretical and observational.

\end{enumerate}

\section*{Acknowledgements}

The observations have been partially supported by the OPTICON project (observing proposals 2010B/01, 2011A/014 and 2012A/015), which is funded by the European Commission under the Seventh Framework Programme (FP7). Part of the observations have been taken under service mode (proposal AO171) and the authors gratefully acknowledge the help of the AAO support astronomers  This research is partially supported by the Spanish Government Ministerio de Econom{\'i}a y Competitividad under grants  FJCI-2014-23001 and AYA2015-68012-C2-2-P (MINECO/FEDER). This research has made use of the Simbad database, operated at CDS, Strasbourg (France). This publication also made use of data products from the Two Micron All Sky Survey, which is a joint project of the University of Massachusetts and the Infrared Processing and Analysis Center/California Institute of Technology, funded by the National Aeronautics and Space Administration and the National Science Foundation.




\bibliographystyle{mnras}
\bibliography{tscale_RSGs} 

\begin{thebibliography}{}
\makeatletter
\relax
\def\mn@urlcharsother{\let\do\@makeother \do\$\do\&\do\#\do\^\do\_\do\%\do\~}
\def\mn@doi{\begingroup\mn@urlcharsother \@ifnextchar [ {\mn@doi@}
  {\mn@doi@[]}}
\def\mn@doi@[#1]#2{\def\@tempa{#1}\ifx\@tempa\@empty \href
  {http://dx.doi.org/#2} {doi:#2}\else \href {http://dx.doi.org/#2} {#1}\fi
  \endgroup}
\def\mn@eprint#1#2{\mn@eprint@#1:#2::\@nil}
\def\mn@eprint@arXiv#1{\href {http://arxiv.org/abs/#1} {{\tt arXiv:#1}}}
\def\mn@eprint@dblp#1{\href {http://dblp.uni-trier.de/rec/bibtex/#1.xml}
  {dblp:#1}}
\def\mn@eprint@#1:#2:#3:#4\@nil{\def\@tempa {#1}\def\@tempb {#2}\def\@tempc
  {#3}\ifx \@tempc \@empty \let \@tempc \@tempb \let \@tempb \@tempa \fi \ifx
  \@tempb \@empty \def\@tempb {arXiv}\fi \@ifundefined
  {mn@eprint@\@tempb}{\@tempb:\@tempc}{\expandafter \expandafter \csname
  mn@eprint@\@tempb\endcsname \expandafter{\@tempc}}}

\bibitem[\protect\citeauthoryear{{Asplund}, {Grevesse}  \& {Sauval}}{{Asplund}
  et~al.}{2005}]{asp05}
{Asplund} M.,  {Grevesse} N.,   {Sauval} A.~J.,  2005, in {Barnes} III T.~G.,
  {Bash} F.~N.,  eds,  Astronomical Society of the Pacific Conference Series
  Vol. 336, Cosmic Abundances as Records of Stellar Evolution and
  Nucleosynthesis. p.~25

\bibitem[\protect\citeauthoryear{{Barklem}, {Piskunov}  \& {O'Mara}}{{Barklem}
  et~al.}{2000}]{bar00}
{Barklem} P.~S.,  {Piskunov} N.,   {O'Mara} B.~J.,  2000, \mn@doi [\aaps]
  {10.1051/aas:2000167}, \href
  {http://adsabs.harvard.edu/abs/2000A%26AS..142..467B} {142, 467}

\bibitem[\protect\citeauthoryear{{Beasor} \& {Davies}}{{Beasor} \&
  {Davies}}{2016}]{bea2016}
{Beasor} E.~R.,  {Davies} B.,  2016, \mn@doi [\mnras] {10.1093/mnras/stw2054},
  \href {http://adsabs.harvard.edu/abs/2016MNRAS.463.1269B} {463, 1269}

\bibitem[\protect\citeauthoryear{{Bergemann}, {Kudritzki}, {Plez}, {Davies},
  {Lind}  \& {Gazak}}{{Bergemann} et~al.}{2012}]{ber12}
{Bergemann} M.,  {Kudritzki} R.-P.,  {Plez} B.,  {Davies} B.,  {Lind} K.,
  {Gazak} Z.,  2012, \mn@doi [\apj] {10.1088/0004-637X/751/2/156}, \href
  {http://adsabs.harvard.edu/abs/2012ApJ...751..156B} {751, 156}

\bibitem[\protect\citeauthoryear{{Bergemann}, {Kudritzki}, {W{\"u}rl}, {Plez},
  {Davies}  \& {Gazak}}{{Bergemann} et~al.}{2013}]{ber13}
{Bergemann} M.,  {Kudritzki} R.-P.,  {W{\"u}rl} M.,  {Plez} B.,  {Davies} B.,
  {Gazak} Z.,  2013, \mn@doi [\apj] {10.1088/0004-637X/764/2/115}, \href
  {http://adsabs.harvard.edu/abs/2013ApJ...764..115B} {764, 115}

\bibitem[\protect\citeauthoryear{{Bergemann}, {Kudritzki}, {Gazak}, {Davies}
  \& {Plez}}{{Bergemann} et~al.}{2015}]{ber15}
{Bergemann} M.,  {Kudritzki} R.-P.,  {Gazak} Z.,  {Davies} B.,   {Plez} B.,
  2015, \mn@doi [\apj] {10.1088/0004-637X/804/2/113}, \href
  {http://adsabs.harvard.edu/abs/2015ApJ...804..113B} {804, 113}

\bibitem[\protect\citeauthoryear{{Bessell} \& {Wood}}{{Bessell} \&
  {Wood}}{1984}]{bes1984}
{Bessell} M.~S.,  {Wood} P.~R.,  1984, \mn@doi [\pasp] {10.1086/131328}, \href
  {http://adsabs.harvard.edu/abs/1984PASP...96..247B} {96, 247}

\bibitem[\protect\citeauthoryear{{Brott} et~al.,}{{Brott}
  et~al.}{2011}]{bro2011}
{Brott} I.,  et~al., 2011, \mn@doi [\aap] {10.1051/0004-6361/201016113}, \href
  {http://adsabs.harvard.edu/abs/2011A%26A...530A.115B} {530, A115}

\bibitem[\protect\citeauthoryear{{Davies}, {Kudritzki}  \& {Figer}}{{Davies}
  et~al.}{2010}]{dav2010}
{Davies} B.,  {Kudritzki} R.-P.,   {Figer} D.~F.,  2010, \mn@doi [\mnras]
  {10.1111/j.1365-2966.2010.16965.x}, \href
  {http://adsabs.harvard.edu/abs/2010MNRAS.407.1203D} {407, 1203}

\bibitem[\protect\citeauthoryear{{Davies} et~al.,}{{Davies}
  et~al.}{2013}]{dav2013}
{Davies} B.,  et~al., 2013, \mn@doi [\apj] {10.1088/0004-637X/767/1/3}, \href
  {http://adsabs.harvard.edu/abs/2013ApJ...767....3D} {767, 3}

\bibitem[\protect\citeauthoryear{{Davies}, {Kudritzki}, {Gazak}, {Plez},
  {Bergemann}, {Evans}  \& {Patrick}}{{Davies} et~al.}{2015}]{dav2015}
{Davies} B.,  {Kudritzki} R.-P.,  {Gazak} Z.,  {Plez} B.,  {Bergemann} M.,
  {Evans} C.,   {Patrick} L.,  2015, \mn@doi [\apj]
  {10.1088/0004-637X/806/1/21}, \href
  {http://adsabs.harvard.edu/abs/2015ApJ...806...21D} {806, 21}

\bibitem[\protect\citeauthoryear{{Dorda}, {Gonz{\'a}lez-Fern{\'a}ndez}  \&
  {Negueruela}}{{Dorda} et~al.}{2016b}]{dor16b}
{Dorda} R.,  {Gonz{\'a}lez-Fern{\'a}ndez} C.,   {Negueruela} I.,  2016b, \aap

\bibitem[\protect\citeauthoryear{{Dorda}, {Negueruela},
  {Gonz{\'a}lez-Fern{\'a}ndez}  \& {Tabernero}}{{Dorda} et~al.}{2016a}]{dor16a}
{Dorda} R.,  {Negueruela} I.,  {Gonz{\'a}lez-Fern{\'a}ndez} C.,   {Tabernero}
  H.~M.,  2016a, \aap

\bibitem[\protect\citeauthoryear{{Drout}, {Massey}  \& {Meynet}}{{Drout}
  et~al.}{2012}]{dro2012}
{Drout} M.~R.,  {Massey} P.,   {Meynet} G.,  2012, \mn@doi [\apj]
  {10.1088/0004-637X/750/2/97}, \href
  {http://adsabs.harvard.edu/abs/2012ApJ...750...97D} {750, 97}

\bibitem[\protect\citeauthoryear{{Ekstr{\"o}m} et~al.,}{{Ekstr{\"o}m}
  et~al.}{2012}]{eks2012}
{Ekstr{\"o}m} S.,  et~al., 2012, \mn@doi [\aap] {10.1051/0004-6361/201117751},
  \href {http://adsabs.harvard.edu/abs/2012A%26A...537A.146E} {537, A146}

\bibitem[\protect\citeauthoryear{{Ekstr{\"o}m}, {Georgy}, {Meynet}, {Groh}  \&
  {Granada}}{{Ekstr{\"o}m} et~al.}{2013}]{eks2013}
{Ekstr{\"o}m} S.,  {Georgy} C.,  {Meynet} G.,  {Groh} J.,   {Granada} A.,
  2013, in {Kervella} P.,  {Le Bertre} T.,   {Perrin} G.,  eds,  EAS
  Publications Series Vol. 60, EAS Publications Series. pp 31--41 (\mn@eprint
  {arXiv} {1303.1629}), \mn@doi{10.1051/eas/1360003}

\bibitem[\protect\citeauthoryear{{Elias}, {Frogel}  \& {Humphreys}}{{Elias}
  et~al.}{1985}]{eli1985}
{Elias} J.~H.,  {Frogel} J.~A.,   {Humphreys} R.~M.,  1985, \mn@doi [\apjs]
  {10.1086/190997}, \href {http://adsabs.harvard.edu/abs/1985ApJS...57...91E}
  {57, 91}

\bibitem[\protect\citeauthoryear{Freedman \& Diaconis}{Freedman \&
  Diaconis}{1981}]{fdr81}
Freedman D.,  Diaconis P.,  1981, \mn@doi [Zeitschrift f{\"u}r
  Wahrscheinlichkeitstheorie und Verwandte Gebiete] {10.1007/BF01025868}, 57,
  453

\bibitem[\protect\citeauthoryear{{Garc{\'{\i}}a-Hern{\'a}ndez},
  {Garc{\'{\i}}a-Lario}, {Plez}, {Manchado}, {D'Antona}, {Lub}  \&
  {Habing}}{{Garc{\'{\i}}a-Hern{\'a}ndez} et~al.}{2007}]{gher07}
{Garc{\'{\i}}a-Hern{\'a}ndez} D.~A.,  {Garc{\'{\i}}a-Lario} P.,  {Plez} B.,
  {Manchado} A.,  {D'Antona} F.,  {Lub} J.,   {Habing} H.,  2007, \mn@doi
  [\aap] {10.1051/0004-6361:20065785}, \href
  {http://adsabs.harvard.edu/abs/2007A%26A...462..711G} {462, 711}

\bibitem[\protect\citeauthoryear{{Gazak}, {Davies}, {Kudritzki}, {Bergemann}
  \& {Plez}}{{Gazak} et~al.}{2014}]{gaz2014}
{Gazak} J.~Z.,  {Davies} B.,  {Kudritzki} R.,  {Bergemann} M.,   {Plez} B.,
  2014, \mn@doi [\apj] {10.1088/0004-637X/788/1/58}, \href
  {http://adsabs.harvard.edu/abs/2014ApJ...788...58G} {788, 58}

\bibitem[\protect\citeauthoryear{{Gazak} et~al.,}{{Gazak}
  et~al.}{2015}]{gaz2015}
{Gazak} J.~Z.,  et~al., 2015, \mn@doi [\apj] {10.1088/0004-637X/805/2/182},
  \href {http://adsabs.harvard.edu/abs/2015ApJ...805..182G} {805, 182}

\bibitem[\protect\citeauthoryear{{Georgy} et~al.,}{{Georgy}
  et~al.}{2013}]{geo2013}
{Georgy} C.,  et~al., 2013, \mn@doi [\aap] {10.1051/0004-6361/201322178}, \href
  {http://adsabs.harvard.edu/abs/2013A%26A...558A.103G} {558, A103}

\bibitem[\protect\citeauthoryear{{Gonz{\'a}lez-Fern{\'a}ndez}, {Dorda},
  {Negueruela}  \& {Marco}}{{Gonz{\'a}lez-Fern{\'a}ndez} et~al.}{2015}]{gon15}
{Gonz{\'a}lez-Fern{\'a}ndez} C.,  {Dorda} R.,  {Negueruela} I.,   {Marco} A.,
  2015, \mn@doi [\aap] {10.1051/0004-6361/201425362}, \href
  {http://adsabs.harvard.edu/abs/2015A%26A...578A...3G} {578, A3}

\bibitem[\protect\citeauthoryear{{Graczyk} et~al.,}{{Graczyk}
  et~al.}{2014}]{gra2014}
{Graczyk} D.,  et~al., 2014, \mn@doi [\apj] {10.1088/0004-637X/780/1/59}, \href
  {http://adsabs.harvard.edu/abs/2014ApJ...780...59G} {780, 59}

\bibitem[\protect\citeauthoryear{{Gray} \& {Corbally}}{{Gray} \&
  {Corbally}}{1994}]{graco94}
{Gray} R.~O.,  {Corbally} C.~J.,  1994, \mn@doi [\aj] {10.1086/116893}, \href
  {http://adsabs.harvard.edu/abs/1994AJ....107..742G} {107, 742}

\bibitem[\protect\citeauthoryear{{Gustafsson}, {Edvardsson}, {Eriksson},
  {J{\o}rgensen}, {Nordlund}  \& {Plez}}{{Gustafsson} et~al.}{2008}]{gus08}
{Gustafsson} B.,  {Edvardsson} B.,  {Eriksson} K.,  {J{\o}rgensen} U.~G.,
  {Nordlund} {\AA}.,   {Plez} B.,  2008, \mn@doi [\aap]
  {10.1051/0004-6361:200809724}, \href
  {http://adsabs.harvard.edu/abs/2008A%26A...486..951G} {486, 951}

\bibitem[\protect\citeauthoryear{{Heiter} \& {Eriksson}}{{Heiter} \&
  {Eriksson}}{2006}]{hei06}
{Heiter} U.,  {Eriksson} K.,  2006, \mn@doi [\aap]
  {10.1051/0004-6361:20064925}, \href
  {http://adsabs.harvard.edu/abs/2006A%26A...452.1039H} {452, 1039}

\bibitem[\protect\citeauthoryear{{Humphreys}}{{Humphreys}}{1979}]{hum1979b}
{Humphreys} R.~M.,  1979, \mn@doi [\apjs] {10.1086/190578}, \href
  {http://adsabs.harvard.edu/abs/1979ApJS...39..389H} {39, 389}

\bibitem[\protect\citeauthoryear{{Humphreys} \& {McElroy}}{{Humphreys} \&
  {McElroy}}{1984}]{hum1984}
{Humphreys} R.~M.,  {McElroy} D.~B.,  1984, \mn@doi [\apj] {10.1086/162439},
  \href {http://adsabs.harvard.edu/abs/1984ApJ...284..565H} {284, 565}

\bibitem[\protect\citeauthoryear{{Kupka}, {Ryabchikova}, {Piskunov}, {Stempels}
   \& {Weiss}}{{Kupka} et~al.}{2000}]{kup00}
{Kupka} F.~G.,  {Ryabchikova} T.~A.,  {Piskunov} N.~E.,  {Stempels} H.~C.,
  {Weiss} W.~W.,  2000, Baltic Astronomy, \href
  {http://adsabs.harvard.edu/abs/2000BaltA...9..590K} {9, 590}

\bibitem[\protect\citeauthoryear{{Lee}}{{Lee}}{1970}]{lee1970}
{Lee} T.~A.,  1970, \mn@doi [\apj] {10.1086/150648}, \href
  {http://adsabs.harvard.edu/abs/1970ApJ...162..217L} {162, 217}

\bibitem[\protect\citeauthoryear{{Levesque}}{{Levesque}}{2013}]{lev2013}
{Levesque} E.~M.,  2013, in {Kervella} P.,  {Le Bertre} T.,   {Perrin} G.,
  eds,  EAS Publications Series Vol. 60, EAS Publications Series. pp 269--277
  (\mn@eprint {arXiv} {1302.0822}), \mn@doi{10.1051/eas/1360031}

\bibitem[\protect\citeauthoryear{{Levesque} \& {Massey}}{{Levesque} \&
  {Massey}}{2012}]{lev2012}
{Levesque} E.~M.,  {Massey} P.,  2012, \mn@doi [\aj]
  {10.1088/0004-6256/144/1/2}, \href
  {http://adsabs.harvard.edu/abs/2012AJ....144....2L} {144, 2}

\bibitem[\protect\citeauthoryear{{Levesque}, {Massey}, {Olsen}, {Plez},
  {Josselin}, {Maeder}  \& {Meynet}}{{Levesque} et~al.}{2005}]{lev2005}
{Levesque} E.~M.,  {Massey} P.,  {Olsen} K.~A.~G.,  {Plez} B.,  {Josselin} E.,
  {Maeder} A.,   {Meynet} G.,  2005, \mn@doi [\apj] {10.1086/430901}, \href
  {http://adsabs.harvard.edu/abs/2005ApJ...628..973L} {628, 973}

\bibitem[\protect\citeauthoryear{{Levesque}, {Massey}, {Olsen}, {Plez},
  {Meynet}  \& {Maeder}}{{Levesque} et~al.}{2006}]{lev2006}
{Levesque} E.~M.,  {Massey} P.,  {Olsen} K.~A.~G.,  {Plez} B.,  {Meynet} G.,
  {Maeder} A.,  2006, \mn@doi [\apj] {10.1086/504417}, \href
  {http://adsabs.harvard.edu/abs/2006ApJ...645.1102L} {645, 1102}

\bibitem[\protect\citeauthoryear{{Levesque}, {Massey}, {Olsen}  \&
  {Plez}}{{Levesque} et~al.}{2007}]{lev2007}
{Levesque} E.~M.,  {Massey} P.,  {Olsen} K.~A.~G.,   {Plez} B.,  2007, \mn@doi
  [\apj] {10.1086/520797}, \href
  {http://adsabs.harvard.edu/abs/2007ApJ...667..202L} {667, 202}

\bibitem[\protect\citeauthoryear{{Massey}}{{Massey}}{2002}]{mas2002}
{Massey} P.,  2002, \mn@doi [\apjs] {10.1086/338286}, \href
  {http://adsabs.harvard.edu/abs/2002ApJS..141...81M} {141, 81}

\bibitem[\protect\citeauthoryear{{Massey} \& {Olsen}}{{Massey} \&
  {Olsen}}{2003}]{mas2003b}
{Massey} P.,  {Olsen} K.~A.~G.,  2003, \mn@doi [\aj] {10.1086/379558}, \href
  {http://adsabs.harvard.edu/abs/2003AJ....126.2867M} {126, 2867}

\bibitem[\protect\citeauthoryear{{M{\'e}sz{\'a}ros} et~al.,}{{M{\'e}sz{\'a}ros}
  et~al.}{2012}]{mes12}
{M{\'e}sz{\'a}ros} S.,  et~al., 2012, \mn@doi [\aj]
  {10.1088/0004-6256/144/4/120}, \href
  {http://adsabs.harvard.edu/abs/2012AJ....144..120M} {144, 120}

\bibitem[\protect\citeauthoryear{{Metropolis}, {Rosenbluth}, {Rosenbluth},
  {Teller}  \& {Teller}}{{Metropolis} et~al.}{1953}]{met53}
{Metropolis} N.,  {Rosenbluth} A.~W.,  {Rosenbluth} M.~N.,  {Teller} A.~H.,
  {Teller} E.,  1953, \mn@doi [\jcp] {10.1063/1.1699114}, \href
  {http://adsabs.harvard.edu/abs/1953JChPh..21.1087M} {21, 1087}

\bibitem[\protect\citeauthoryear{{Meynet} \& {Maeder}}{{Meynet} \&
  {Maeder}}{2000}]{mey2000}
{Meynet} G.,  {Maeder} A.,  2000, \aap, \href
  {http://adsabs.harvard.edu/abs/2000A%26A...361..101M} {361, 101}

\bibitem[\protect\citeauthoryear{{Patrick}, {Evans}, {Davies}, {Kudritzki},
  {Gazak}, {Bergemann}, {Plez}  \& {Ferguson}}{{Patrick}
  et~al.}{2015}]{pat2015}
{Patrick} L.~R.,  {Evans} C.~J.,  {Davies} B.,  {Kudritzki} R.-P.,  {Gazak}
  J.~Z.,  {Bergemann} M.,  {Plez} B.,   {Ferguson} A.~M.~N.,  2015, \mn@doi
  [\apj] {10.1088/0004-637X/803/1/14}, \href
  {http://adsabs.harvard.edu/abs/2015ApJ...803...14P} {803, 14}

\bibitem[\protect\citeauthoryear{{Patrick}, {Evans}, {Davies}, {Kudritzki},
  {H{\'e}nault-Brunet}, {Bastian}, {Lapenna}  \& {Bergemann}}{{Patrick}
  et~al.}{2016}]{pat2016}
{Patrick} L.~R.,  {Evans} C.~J.,  {Davies} B.,  {Kudritzki} R.-P.,
  {H{\'e}nault-Brunet} V.,  {Bastian} N.,  {Lapenna} E.,   {Bergemann} M.,
  2016, \mn@doi [\mnras] {10.1093/mnras/stw561}, \href
  {http://adsabs.harvard.edu/abs/2016MNRAS.458.3968P} {458, 3968}

\bibitem[\protect\citeauthoryear{{Piskunov}, {Kupka}, {Ryabchikova}, {Weiss}
  \& {Jeffery}}{{Piskunov} et~al.}{1995}]{pis95}
{Piskunov} N.~E.,  {Kupka} F.,  {Ryabchikova} T.~A.,  {Weiss} W.~W.,
  {Jeffery} C.~S.,  1995, \aaps, \href
  {http://adsabs.harvard.edu/abs/1995A%26AS..112..525P} {112, 525}

\bibitem[\protect\citeauthoryear{{Recio-Blanco}, {Bijaoui}  \& {de
  Laverny}}{{Recio-Blanco} et~al.}{2006}]{rec06}
{Recio-Blanco} A.,  {Bijaoui} A.,   {de Laverny} P.,  2006, \mn@doi [\mnras]
  {10.1111/j.1365-2966.2006.10455.x}, \href
  {http://adsabs.harvard.edu/abs/2006MNRAS.370..141R} {370, 141}

\bibitem[\protect\citeauthoryear{{Ryabchikova}, {Piskunov}, {Kurucz},
  {Stempels}, {Heiter}, {Pakhomov}  \& {Barklem}}{{Ryabchikova}
  et~al.}{2015}]{rya15}
{Ryabchikova} T.,  {Piskunov} N.,  {Kurucz} R.~L.,  {Stempels} H.~C.,  {Heiter}
  U.,  {Pakhomov} Y.,   {Barklem} P.~S.,  2015, \mn@doi [\physscr]
  {10.1088/0031-8949/90/5/054005}, \href
  {http://adsabs.harvard.edu/abs/2015PhyS...90e4005R} {90, 054005}

\bibitem[\protect\citeauthoryear{{Sch{\"o}nrich} \&
  {Bergemann}}{{Sch{\"o}nrich} \& {Bergemann}}{2014}]{sch14}
{Sch{\"o}nrich} R.,  {Bergemann} M.,  2014, \mn@doi [\mnras]
  {10.1093/mnras/stu1072}, \href
  {http://adsabs.harvard.edu/abs/2014MNRAS.443..698S} {443, 698}

\bibitem[\protect\citeauthoryear{{Skrutskie} et~al.,}{{Skrutskie}
  et~al.}{2006}]{skr2006}
{Skrutskie} M.~F.,  et~al., 2006, \mn@doi [\aj] {10.1086/498708}, \href
  {http://adsabs.harvard.edu/abs/2006AJ....131.1163S} {131, 1163}

\bibitem[\protect\citeauthoryear{{Tabernero}, {Montes}  \& {Gonz{\'a}lez
  Hern{\'a}ndez}}{{Tabernero} et~al.}{2012}]{tab12}
{Tabernero} H.~M.,  {Montes} D.,   {Gonz{\'a}lez Hern{\'a}ndez} J.~I.,  2012,
  \mn@doi [\aap] {10.1051/0004-6361/201117506}, \href
  {http://adsabs.harvard.edu/abs/2012A%26A...547A..13T} {547, A13}

\bibitem[\protect\citeauthoryear{{Ting}, {Conroy}, {Rix}  \& {Cargile}}{{Ting}
  et~al.}{2017}]{tin17}
{Ting} Y.-S.,  {Conroy} C.,  {Rix} H.-W.,   {Cargile} P.,  2017, \mn@doi [\apj]
  {10.3847/1538-4357/aa7688}, \href
  {http://adsabs.harvard.edu/abs/2017ApJ...843...32T} {843, 32}

\bibitem[\protect\citeauthoryear{{Ventura}, {D'Antona}  \&
  {Mazzitelli}}{{Ventura} et~al.}{2000}]{ventura00}
{Ventura} P.,  {D'Antona} F.,   {Mazzitelli} I.,  2000, \aap, \href
  {http://adsabs.harvard.edu/abs/2000A%26A...363..605V} {363, 605}

\bibitem[\protect\citeauthoryear{{Walker}}{{Walker}}{2012}]{wal2012}
{Walker} A.~R.,  2012, \mn@doi [\apss] {10.1007/s10509-011-0961-x}, \href
  {http://adsabs.harvard.edu/abs/2012Ap%26SS.341...43W} {341, 43}

\makeatother
\end{thebibliography}




\appendix

\section{Extra material}

\begin{table*}
\centering
\caption{Lines used to derive stellar parameters, we show only predominant features on each spectral window.}
\label{linTab}
\begin{tabular}{ccccccc}
  \hline
  Rank & $\lambda$ interval (\r{A}) & $\lambda_{\rm line}$ (\r{A}) & Element & $\chi_{\rm low}$ (eV) & $\log{\rm gf}$ (dex) & $\log{\gamma_6}$/ABO data \\
  \hline
   1 & $8514.40\pm1.90$ & 8514.09 & \ion{Fe}{i} & 2.20 & -2.23 & 257.246 \\
     &                  & 8515.12 & \ion{Fe}{i} & 3.01 & -2.08 & -7.77   \\
   2 & $8518.30\pm1.50$ & 8518.03 & \ion{Ti}{i} & 2.13 & -0.94 & -7.80   \\
     &                  & 8518.35 & \ion{Ti}{i} & 1.87 & -1.05 & -7.75   \\
   3 & $8582.35\pm1.35$ & 8582.26 & \ion{Fe}{i} & 2.99 & -2.13 & -7.12   \\
   4 & $8611.80\pm0.90$ & 8611.80 & \ion{Fe}{i} & 2.85 & -1.93 & 312.264 \\
   5 & $8679.00\pm2.10$ & 8679.00 & \ion{Fe}{i} & 6.02 & -3.29 & -7.13   \\
     &                  & 8679.63 & \ion{Fe}{i} & 4.97 & -1.29 & -7.52   \\
   6 & $8682.90\pm1.30$ & 8682.98 & \ion{Ti}{i} & 1.05 & -1.79 & 283.246 \\
   7 & $8688.95\pm1.65$ & 8688.62 & \ion{Fe}{i} & 2.18 & -1.21 & 253.245 \\
   8 & $8692.00\pm1.00$ & 8692.33 & \ion{Ti}{i} & 1.05 & -2.13 & -7.75  \\
   9 & $8711.50\pm3.00$ & 8710.18 & \ion{Mg}{i} & 5.93 & -0.54 & --   \\
     &                  & 8710.39 & \ion{Fe}{i} & 4.91 & -0.54 & -7.51  \\
     &                  & 8712.68 & \ion{Mg}{i} & 5.93 & -1.57 & --   \\
     &                  & 8713.19 & \ion{Fe}{i} & 2.94 & -2.47 & -7.50  \\
  10 & $8730.75\pm0.95$ & 8730.50 & \ion{Ti}{i} & 3.35 & -1.57 & -7.75  \\
  11 & $8735.25\pm1.75$ & 8734.71 & \ion{Ti}{i} & 1.05 & -2.24 & 283.247 \\
     &                  & 8736.02 & \ion{Mg}{i} & 5.94 & -0.73 & --  \\
  12 & $8742.25\pm0.75$ & 8742.45 & \ion{Si}{i} & 5.87 & -0.07 & -7.33 \\
  13 & $8757.20\pm1.60$ & 8757.19 & \ion{Fe}{i} & 4.26 & -2.06 & 310.265\\
  14 & $8792.85\pm1.35$ & 8793.34 & \ion{Fe}{i} & 4.61 & -0.09 & -7.54 \\
  15 & $8806.00\pm2.70$ & 8806.76 & \ion{Mg}{i} & 4.34 & -0.13 & 530.277 \\
     &                  & 8808.17 & \ion{Fe}{i} & 5.01 & -1.08 & -7.49 \\
  16 & $8824.36\pm1.15$ & 8824.22 & \ion{Fe}{i} & 2.20 & -1.54 & 254.245\\
  17 & $8838.75\pm1.25$ & 8838.43 & \ion{Fe}{i} & 2.86 & -2.05 & 310.265\\
  \hline
\end{tabular}
\end{table*}

\begin{figure*}
        \centering
        \includegraphics[trim=1cm 0.2cm 1.6cm 1.2cm,clip,width=14.10cm]{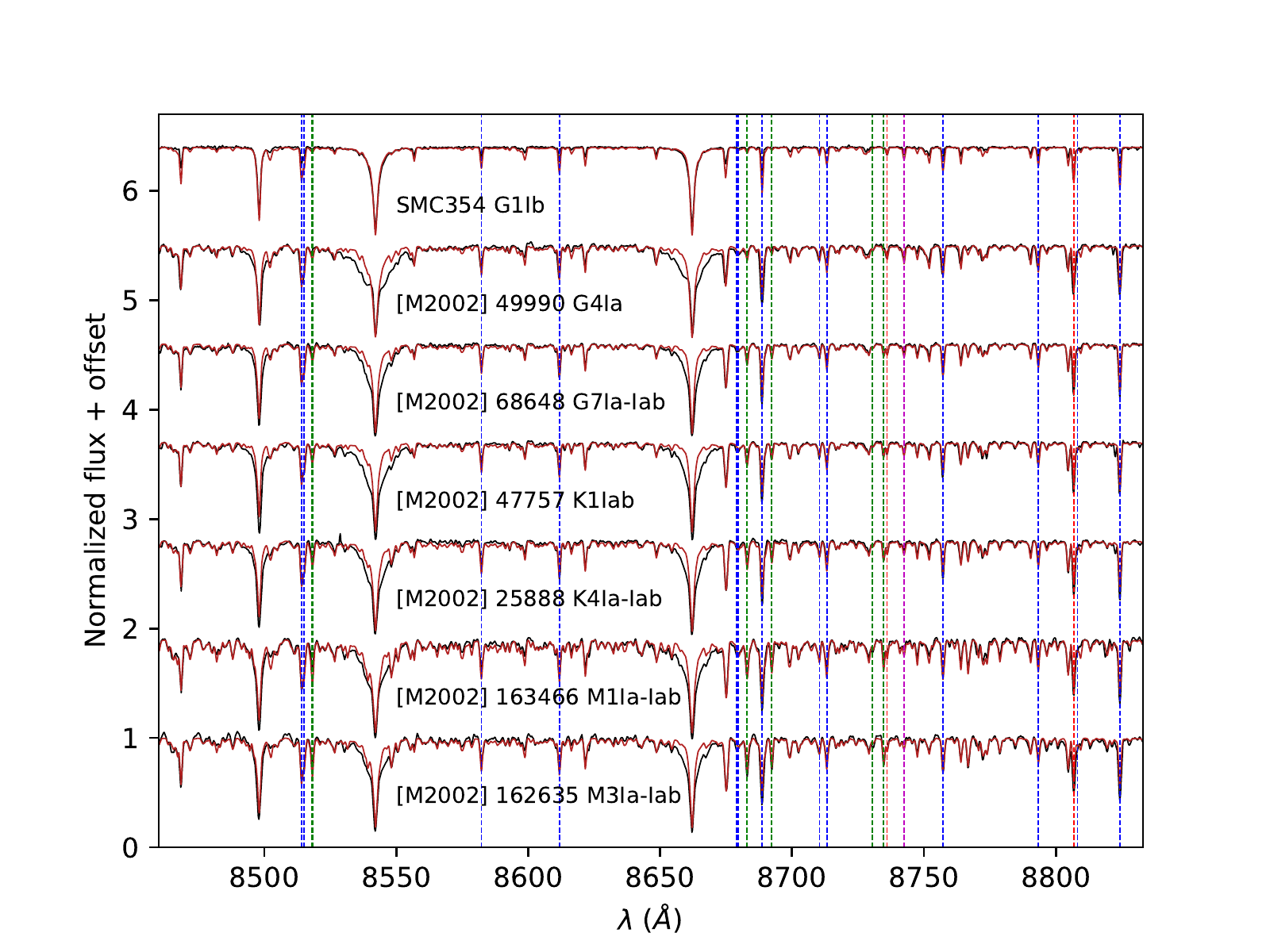}
        \includegraphics[trim=1cm 0.2cm 1.6cm 1.2cm,clip,width=14.10cm]{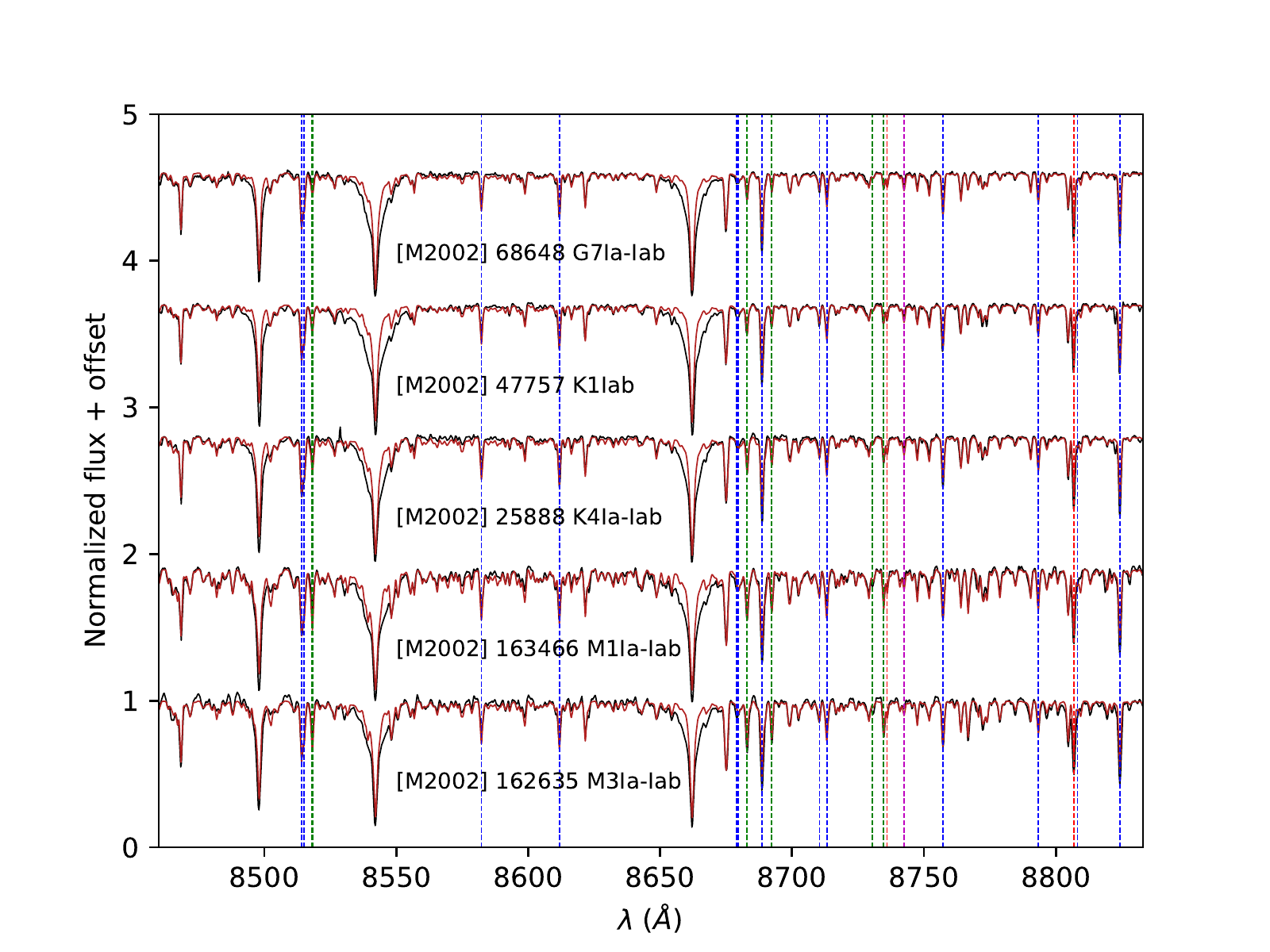}
		\caption{Best-fit synthetic spectra (red line) against observations (black line) for our two synthetic grids: {\bf  Upper pannel (\ref{fitsyn}a)}  (KURUCZ) and {\bf Lower pannel (\ref{fitsyn}b)} (MARCS). Dashed lines refere to the diagnostic lines (see Tab.~\ref{linTab}) we employed in our calculations: Mg (red), Si (magenta), Ti (green), and Fe (blue).} 
                \label{fitsyn}
\end{figure*}

\begin{figure*}
        \centering
        \includegraphics[trim=0.3cm 0.2cm 1.4cm 1.2cm,clip,width=8.5cm]{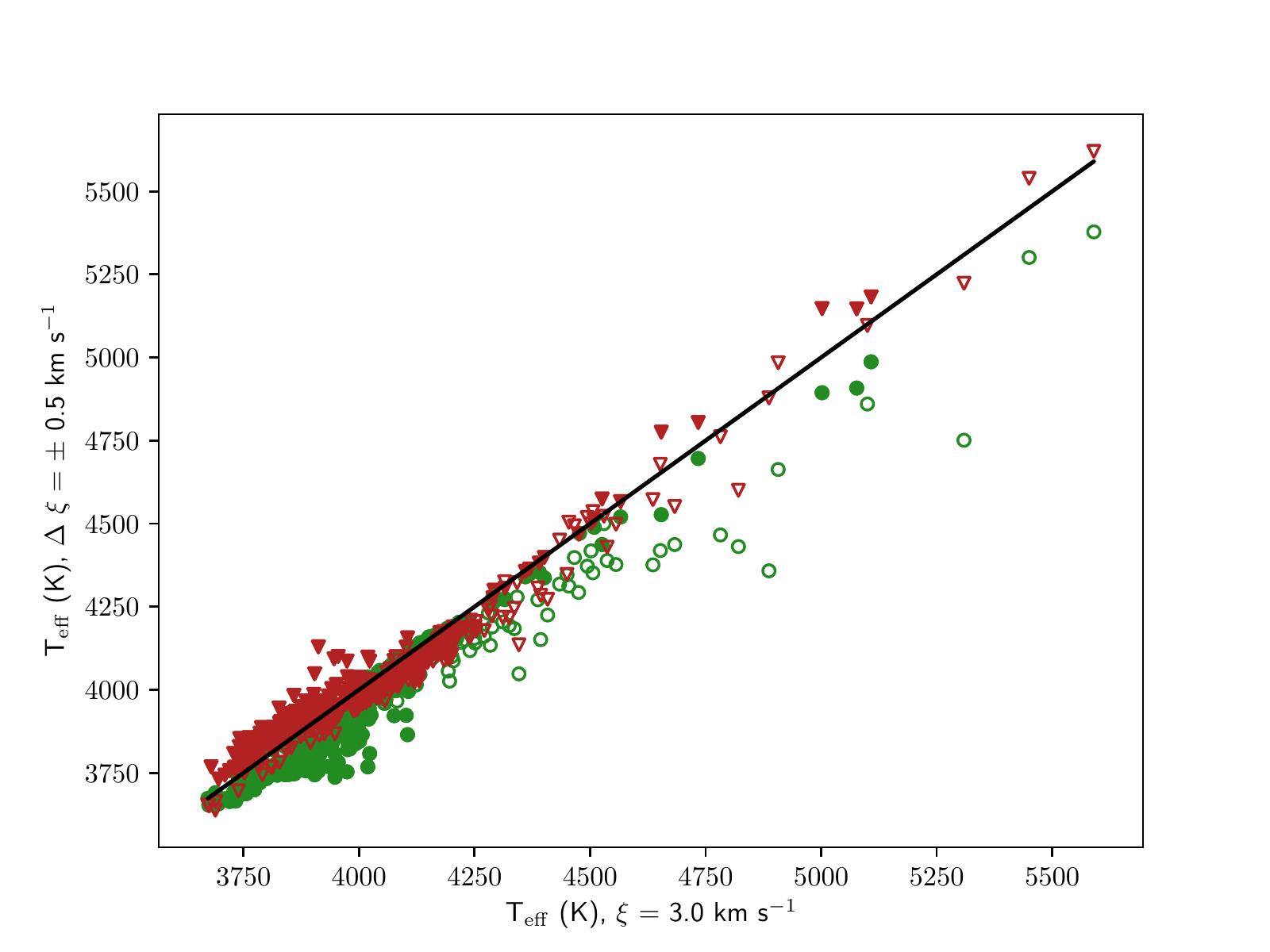}
        \includegraphics[trim=0.3cm 0.2cm 1.4cm 1.2cm,clip,width=8.5cm]{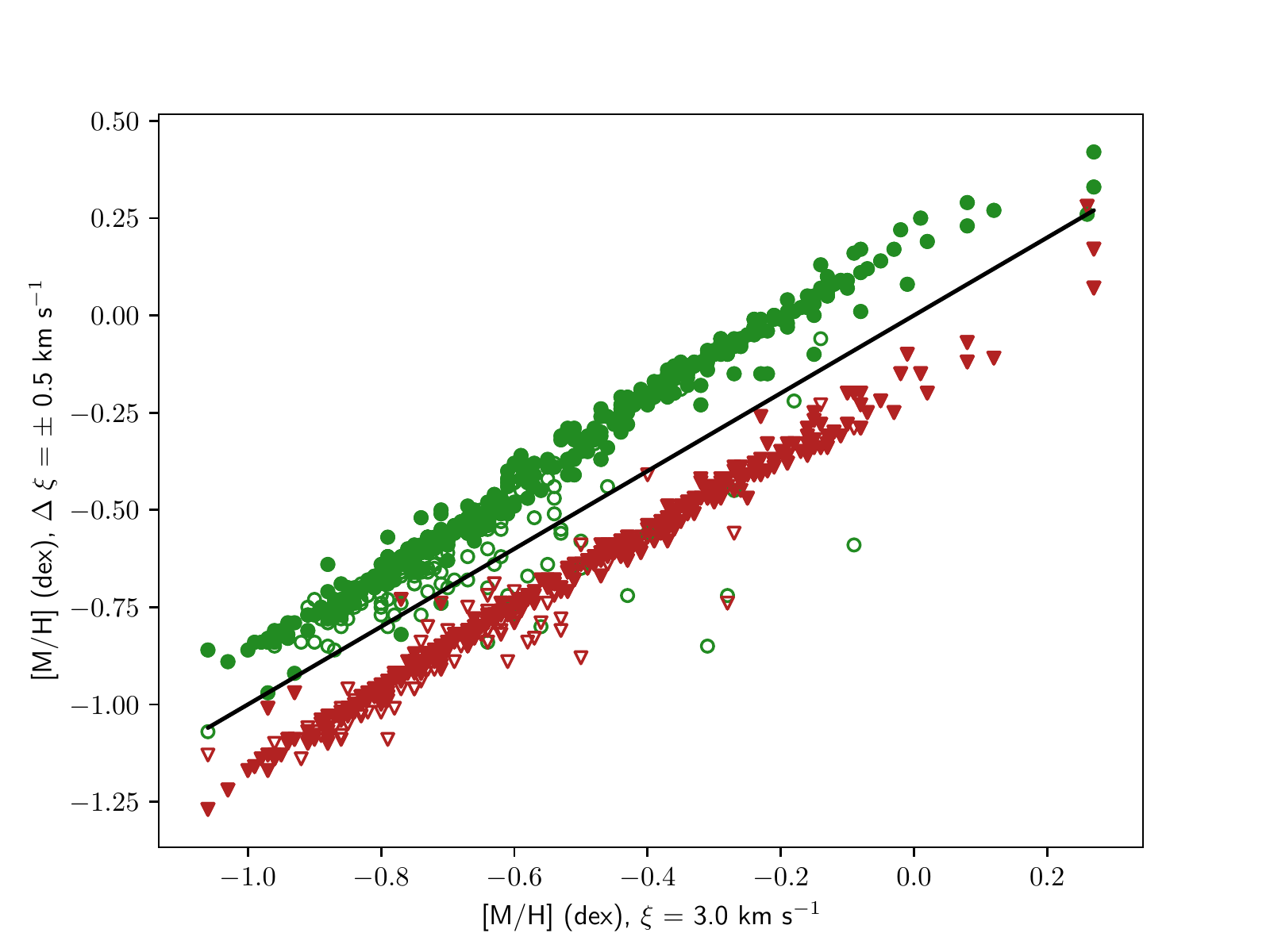}
                \caption{$T_{\rm eff}$ and [M/H] calculated with two different microturbulent velocities ($\xi$~$=$~3~$\pm$~0.5 km~s$^{-1}$) using KURUCZ synthetic spectra. Red triangles correspond to $\xi$~$=$~2.5~km~s$^{-1}$, whereas green circules correspond to $\xi$~$=$~3.5~km~s$^{-1}$. Open simbols represent stars with  uncertainities on $T_{\rm eff}$  of 200~K or more.}
                \label{xiKUR}
\end{figure*}

\begin{figure*}
        \centering
        \includegraphics[trim=0.5cm 0.2cm 1.4cm 1.2cm,clip,width=8.5cm]{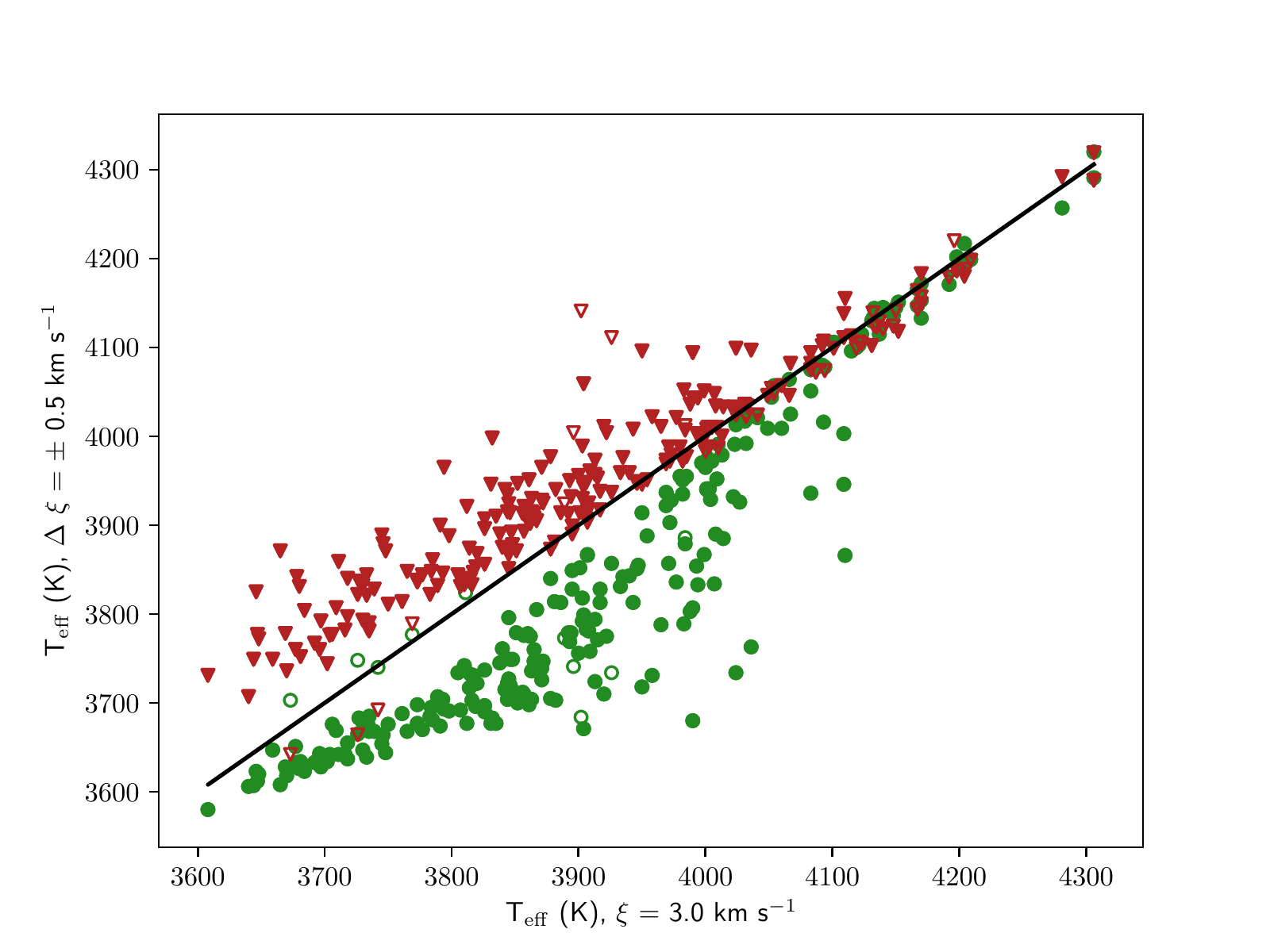}
        \includegraphics[trim=0.5cm 0.2cm 1.4cm 1.2cm,clip,width=8.5cm]{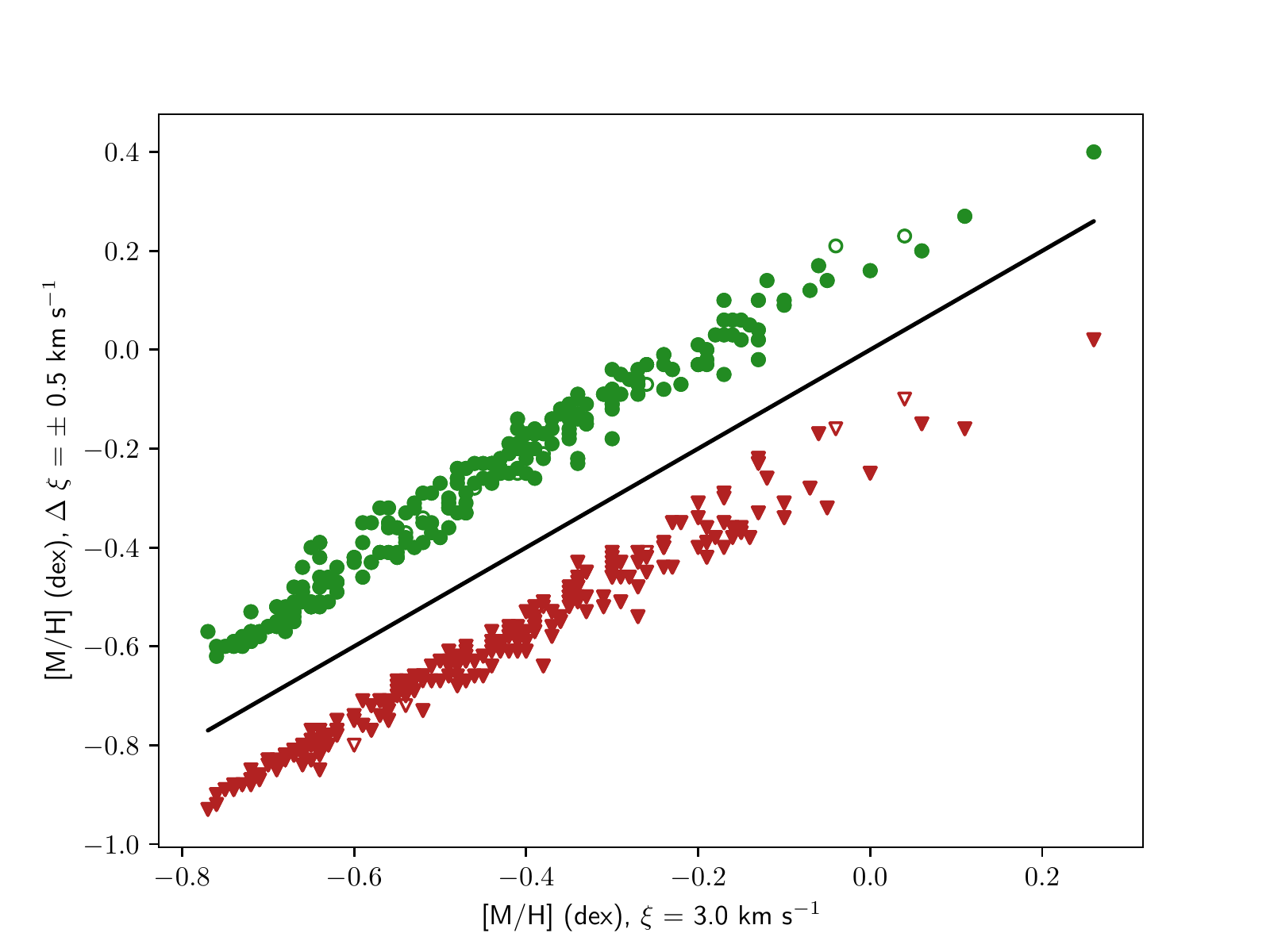}
                \caption{Same as Fig.~\ref{xiKUR} but for MARCS synthetic spectra calculations.}
                \label{xiMAR}
\end{figure*}

\onecolumn
\begin{longtable}[h]{c c | c c | c c | c c | c c | c c c}
\caption{A list of CSGs for which we derived stellar atmospheric parameters ($T_{\rm eff}$ and [M/H]). The SpT and the LC are taken from \citetalias{gon15}, while the $M_{\rm bol}$ correspond to those from \citetalias{dor16a}. Stars tagged as variable are shown in different rows, one for each epoch (Variability is indicated at the column Var.). The epochs are: A=2010, B=2011, C=2012, D=2013.} \label{parGTAB} \\
\noalign{\smallskip}
\hline\hline
\noalign{\smallskip}
ID&Cloud&RA&DEC&Var.&Epochs&$T_{\rm eff}$&$\Delta T_{\rm eff}$&[M/H]&$\Delta$[M/H]&SpT&LC&$M_{\mathrm{bol}}$\\
&&J2000&J2000&&&(K)&(K)&(dex)&(dex)&&&(mag)\\
\noalign{\smallskip}
\hline
\noalign{\smallskip}
\endfirsthead
\caption{continued.}\\
\noalign{\smallskip}
ID&Cloud&RA&DEC&Var.&Epochs&$T_{\rm eff}$&$\Delta T_{\rm eff}$&[M/H]&$\Delta$[M/H]&SpT&LC&$M_{\mathrm{bol}}$\\
&&J2000&J2000&&&(K)&(K)&(dex)&(dex)&&&(mag)\\
\noalign{\smallskip}
\hline
\noalign{\smallskip}
\endhead
8324&SMC&00:47:16.86&-73:08:08.59&yes&B&4121&119&-0.94&0.14&K0&Ia-Iab&-6.88\\
8324&SMC&00:47:16.86&-73:08:08.59&yes&A&4237&307&-0.82&0.37&G7&Iab&-6.88\\
8324&SMC&00:47:16.86&-73:08:08.59&yes&C&4201&174&-0.79&0.19&G7&Iab&-6.88\\
\hline
8367&SMC&00:47:18.14&-73:10:39.32&yes&B&4011&77&-0.82&0.08&K2&Ia-Iab&-7.23\\
8367&SMC&00:47:18.14&-73:10:39.32&yes&A&3981&210&-0.73&0.19&K3.5&Ia&-7.23\\
8367&SMC&00:47:18.14&-73:10:39.32&yes&C&4139&123&-0.60&0.11&G8&Iab&-7.23\\
\hline
8930&SMC&00:47:36.89&-73:04:44.25&yes&B&4076&66&-0.59&0.07&K2.5&Ia-Iab&-7.82\\
8930&SMC&00:47:36.89&-73:04:44.25&yes&A&3778&209&-0.77&0.17&M3&Iab&-7.82\\
\hline
\end{longtable}

\bsp	
\label{lastpage}
\end{document}